\documentclass[
aps,
twocolumn,
prc,
amsmath,
amssymb,
superscriptaddress
]{revtex4-2}

\usepackage{graphicx}
\usepackage{dcolumn}
\usepackage{bm}
\usepackage{subfigure}
\usepackage{hyperref}
\usepackage{amsmath}
\usepackage{hyphenat}
\usepackage{amsfonts}
\usepackage{booktabs}
\usepackage{multirow}
\usepackage{afterpage}
\usepackage[utf8]{inputenc} 
\usepackage{natbib}
\usepackage[usenames]{color}
\hypersetup{bookmarks=true, colorlinks=true,linktoc=page, linkcolor=blue, citecolor=blue, urlcolor=blue }

\begin{document}

\title{Exploring total fission cross sections of neutron- and proton-induced reactions of exotic nuclei at relativistic energies using the INCL-ABLA++ models}

\author{J.~L.~Rodr\'{i}guez-S\'{a}nchez}
\email[Corresponding author: ]{j.l.rodriguez.sanchez@udc.es}
\affiliation{CITENI, Campus Industrial de Ferrol, Universidade da Coru\~{n}a, E-15403 Ferrol, Spain}
\affiliation{IGFAE, Universidade de Santiago de Compostela, E-15782 Santiago de Compostela, Spain}

\author{A.~Gra\~{n}a-Gonz\'{a}lez}
\affiliation{CITENI, Campus Industrial de Ferrol, Universidade da Coru\~{n}a, E-15403 Ferrol, Spain}

\author{J.-C.~David}
\affiliation{IRFU, CEA, Universit\'{e} Paris-Saclay, F-91191 Gif-sur-Yvette, France}

\author{G.~Garc\'{i}a-Jim\'{e}nez}
\affiliation{Lawrence Berkeley National Laboratory, CA-94720 Berkeley, USA}

\author{J.~Hirtz}
\affiliation{IRFU, CEA, Universit\'{e} Paris-Saclay, F-91191 Gif-sur-Yvette, France}
\affiliation{Physics Institute, University of Bern, Sidlerstrasse 5, 3012 Bern, Switzerland}

\author{A.~Keli\'{c}-Heil}
\affiliation{GSI-Helmholtzzentrum f\"{u}r Schwerionenforschung GmbH, D-64291 Darmstadt, Germany}

\date{\today}

\begin{abstract}
Spallation-induced fission reactions, in combination with state-of-the-art dynamical calculations, provide a robust framework for studying the nuclear dissipation mechanism in nuclear matter over a wide range of deformations. Experimental datasets of neutron- and proton-induced fission, accumulated over the last decades, cover a wide range of nuclei from the pre-actinide to the actinide region. In this work, these datasets are used to benchmark the dynamical Liège intranuclear cascade model INCL++ integrated with the ABLA++ de-excitation code, which has been improved by updating the calculation of particle separation energies to the atomic mass evaluation (AME2020) and incorporating phenomenological corrections to enhance the description of fission-barrier heights. The results show good agreement with the available experimental data of total fission cross sections induced by protons and neutrons, thereby confirming the applicability and predictive capability of these models. Moreover, the improved framework is used to investigate the fission cross sections of exotic heavy fissile nuclei, far from the valley of stability, with masses $A > 155$.
\end{abstract}

\maketitle

\section{Introduction}
Nuclear fission, discovered in 1938 by Meitner, Hahn, Strassmann and Frisch \cite{fission_discovery,Hahn1939}, involves the splitting of an atomic nucleus into two smaller fragments, accompanied by the release of significant energy and particle emission. Fission is a critical tool for studying the nuclear potential energy landscape, tracking its evolution as a function of parameters such as elongation, mass asymmetry, angular momentum, and excitation energy~\cite{Schmidt2018,Andreyev2018}. The process involves the interplay of macroscopic effects, including collective mean-field dynamics, and microscopic contributions such as shell effects and nucleon pairing correlations, which influence both the initial compound nucleus and the final fission fragments~\cite{Scamps2018}. The fission mechanism also plays an important role in phenomena such as nuclear stellar explosions \cite{supernova_Truran} and astrophysical r-process nucleosynthesis~\cite{GW17_rprocess, Siegel_review}. The latter is terminated by fission and contributes to the abundance of medium-mass elements via "fission recycling"~\cite{fission_recycling}.

Soon after the discovery of fission, Bohr and Wheeler~\cite{NBPR56} introduced a macroscopic model of the fissioning nucleus based on the liquid-drop analogy, where fission results from the competition between Coulomb repulsion and surface tension forces. This approach served as the basis for transition-state theories, the so-called standard statistical model, which describe fission dynamics using the level densities of the nucleus at the ground state and at the transition state, often identified as the saddle or scission points. Various extensions of this approach have since been developed, mostly based on stochastic methods. This approach was used by Kramers~\cite{kramers1940} to develop the diffusion model that describes nuclear fission using a small set of collective degrees of freedom interacting with a "thermostat" composed of all other single-particle degrees of freedom. Under these circumstances, the collective-variable dynamics becomes similar to Brownian-particle dynamics, since the collective subsystem energy varies minimally during each interaction with the single-particle subsystem. 

Nowadays the behavior of the fission process can be effectively modeled using stochastic methods based on transport models, such as the Langevin equations~\cite{YAPR275} or the equivalent Fokker-Planck equation (FPE) \cite{FPE_book}. These models are able to describe the distribution function of collective coordinates and their conjugate momenta, as well as their time evolution. A key aspect of this modeling is the dissipation of collective energy as the system evolves from a compact shape to the scission point~\cite{AVKPRC63,AJSPRC96,FAIPRC109}, where the fissioning nucleus splits into the fission fragments. The effects of dissipation and transient time become particularly evident at high excitation energies (above 70~MeV), where the average decay time of the fissioning system is comparable to the transient time, leading to a reduction in the fission probability~\cite{BJPLB553,CSPRL07,YAPRC91}. This is supported by many experimental observables, especially the significantly higher rates for the emission of pre-scission neutrons \cite{Hilsher_pre_scission_n, Gavron_pre_scission_n, Shaw_pre_scission_n_gamma, Dioszegi_pre_scission_n_gamma_61_2000, Dioszegi_pre_scission_n_gamma_63_2000, Lestone1999}, charged particles~\cite{Lestone1991}, and $\gamma$-ray \cite{Shaw_pre_scission_n_gamma, Dioszegi_pre_scission_n_gamma_61_2000, Dioszegi_pre_scission_n_gamma_63_2000, Thoennessen_pre_scission_gamma, Hofman_pre_scission_gamma} than those predicted by the statistical model.

Intermediate-energy fission reactions, such as those induced by spallation reactions at kinetic energies exceeding 70~MeV/nucleon, are well-suited for investigating fission dynamics~\cite{Yassid89_2014,YAPRC91,JLRS14,JLRS16_2} and have significant applications in energy production with accelerator-driven systems~\cite{Maschek2008}, radioactive waste transmutation, and radiation shielding design for accelerator systems and cosmic ray devices~\cite{Koning1998}. Additionally, fission enables the production of exotic nuclei far from the valley of stability~\cite{Loureiro2019} at worldwide nuclear physics facilities, such as RIBF, FRIB, and GSI/FAIR. These applications necessitate the determination of proton and neutron fission cross sections with high accuracy over a broad energy range. Apart from its importance for fundamental studies, fission is also relevant for the synthesis of radioisotopes for medical and scientific use, making it both a fundamental and practical area of nuclear physics research.

Over the past seventy years, a substantial amount of experimental data has been collected for neutron- and proton-induced fission reactions. For example,  Ref.~\cite{Obukhov2001} presents a comprehensive review of $(n,f)$ and $(p,f)$ measurements up to the early $21^{st}$ century, while Ref.~\cite{Prokofiev2001} offers a parametrization of $(p,f)$ cross sections derived from the same experimental systematics. Recent $(p,f)$ experiments have been notably advanced by the work of Kotov and collaborators~\cite{Kotov2006} in direct kinematics, who provided detailed cross-section data in 100~MeV intervals over the energy range of 200~MeV to 1~GeV. These measurements cover a variety of actinides critical for applications, including $^{232}\mathrm{Th}$, $^{233,235,238}\mathrm{U}$, $^{237}\mathrm{Np}$, and $^{239}\mathrm{Pu}$, as well as two pre-actinides, $^{nat}\mathrm{Pb}$ and $^{209}\mathrm{Bi}$. The latter nuclei are of particular interest as they form part of the eutectic system utilized both as a spallation target and coolant in accelerator-driven systems.

Currently, the $n\_TOF$ facility~\cite{Tagliente2024} at CERN plays an important role in $(n,f)$ experiments, allowing for the measurement of fission cross sections from thermal energies up to approximately 1~GeV for targets made of stable nuclei. This is achieved through a high-intensity pulsed neutron beam produced by 20~GeV/c protons coming from the PS accelerator, which impinge onto a lead spallation target. Since the determination of absolute fission cross sections requires simultaneous measurement of both fission events and neutron flux, which involves significant experimental challenges, the current measurements are reported relative to the $^{235}\mathrm{U}(n,f)$ reaction. Absolute cross sections are subsequently derived by normalizing the experimental ratios to an evaluated $^{235}\mathrm{U}(n,f)$ cross section, typically sourced from the JENDL/HE-2007 library~\cite{Watanabe2011} for neutron energies between 30~MeV and 1~GeV. Published $(n,f)$ cross-section datasets up to approximately 1~GeV are available for $^{nat}\mathrm{Pb}$~\cite{Tarrio2011}, $^{209}\mathrm{Bi}$~\cite{Tarrio2011}, $^{232}\mathrm{Th}$~\cite{Tarrio2023}, $^{233}\mathrm{U}$~\cite{Tarrio2023}, $^{234}\mathrm{U}$~\cite{Paradela2010}, $^{235}\mathrm{U}$~\cite{Manna2025}, and $^{237}\mathrm{Np}$~\cite{Paradela2010}.

In this work, we employ the available fission cross-section data for neutron- and proton-induced reactions to benchmark the Li\`{e}ge intranuclear cascade model (INCL++) integrated with the de-excitation ABLA++ model. These codes allow for the description of spallation-induced fission reactions through dynamical approaches that model the interaction of neutrons and protons with the nucleus, followed by a competitive de-excitation stage involving particle emission and fission.

\section{Theoretical framework}
Spallation reactions induced by neutrons and protons are modeled using the latest C++ implementation of the dynamical Li\`{e}ge intranuclear cascade model, INCL++(v6.33.8)~\cite{Mancusi14,PRC91Davide2015,JCD2019,JHPRC101,JLRSPLB851,Demid2024}, coupled to the advanced de-excitation model ABLA++(v24.12)~\cite{JLRSPRC105,JL2023}. Both models leverage Monte Carlo algorithms to simulate reaction dynamics while rigorously enforcing fundamental conservation laws, including energy, momentum, and baryon number at each stage of the reaction~\cite{ABPRC87}. ABLA++ represents a modernized C++ translation of the legacy Fortran-based ABLA07 code~\cite{Abla07}, integrating significant advancements in particle emission decay channels and extending its functionality to include hypernuclei formation. The coupling of both models provides the user with the flexibility to select the reference frame for simulations, enabling the choice between direct and inverse kinematics. For simplicity, these C++ code implementations will hereafter be referred to as the INCL and ABLA++ models, while the legacy Fortran-based version will be referred to as ABLA07.

\subsection{INCL model}
INCL describes the neutron-nucleus and proton-nucleus reactions as a sequence of binary collisions between the projectile and the nucleons (hadrons) present in the target nucleus, mainly in the geometrical overlapping region~\cite{Cugnon1996}. Nucleons move along straight trajectories until they undergo a collision with another nucleon or until they reach the surface, where they may escape. INCL incorporates isospin- and energy-dependent nuclear potentials derived from optical model calculations~\cite{ABPRC87}, along with isospin-dependent pion potentials~\cite{Aoust2006}, which are essential for accurately modeling inelastic nucleon-nucleon interactions that lead to the excitation of baryonic resonances~\cite{JLRS2020res,JLRS2022res}. Recently, INCL has been extended to high energies (up to $\sim$20 GeV) by incorporating new interaction processes, including multi-pion production~\cite{SPASR44,Mancusi2017}, the production of $\eta$ and $\omega$ mesons~\cite{JCDEPJP133}, and the generation of strange particles~\cite{Jason2018,JLRSPRC98,JHPRC101}, such as kaons and hyperons. At the first step of the simulation, the target density profile is prepared assuming independent Woods-Saxon density distributions for protons, neutrons, and $\Lambda$-particles~(only used for hypernuclei)~\cite{ABPRC87,JLRSPRC98,Vries1987}. For the Woods-Saxon density distribution, the radius ($R_0$) and the diffuseness parameter ($a$) are taken from Hartree-Fock-Bogoliubov calculations~\cite{JLRS2017incl}. Additionally, the nucleons are sampled in phase space taking into account the correlations between kinetic energy and radius of the potential well~\cite{Bou02}, such that the relationship is given by the Woods-Saxon distribution. 
\begin{figure}[t!]
\centering
\subfigure{\includegraphics[width=0.49\textwidth,keepaspectratio]{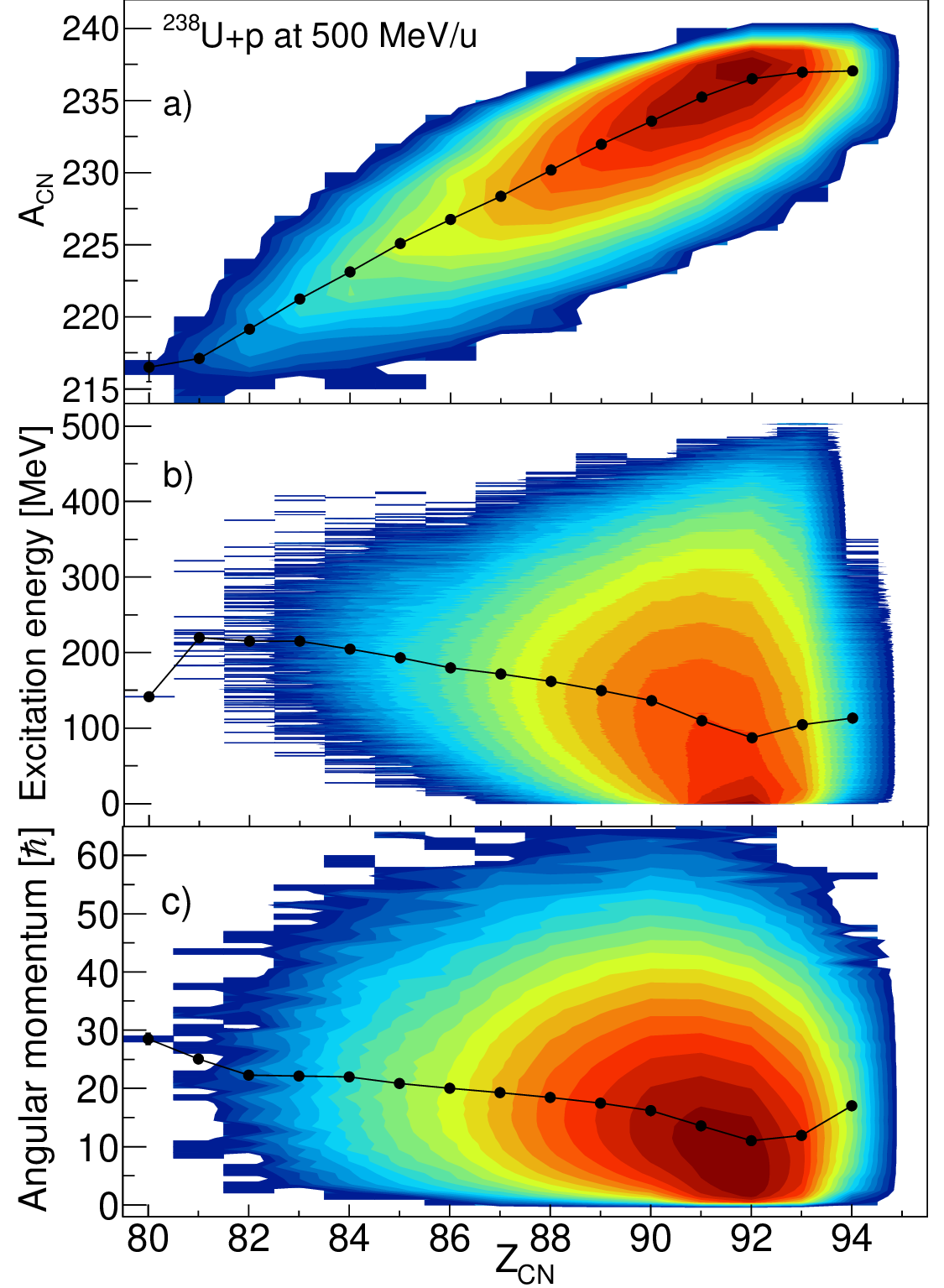}}
\caption{(Color online) Characteristics of compound nuclei produced in the reaction $^{238}$U + p at 500~MeV/nucleon: distributions of masses (a), excitation energy (b), and angular momentum (c) are displayed as a function of the compound nucleus atomic number ($Z_{\mathrm{CN}}$). The dots represent the mean values corresponding to each $Z_{\mathrm{CN}}$.
}
\label{fig1}
\end{figure}
The Pauli blocking also plays a fundamental role in INCL and is treated distinctly for the first collision and the subsequent ones. In the initial collision, a strict blocking is enforced, requiring that both nucleons occupy states outside the Fermi sphere after the interaction. For subsequent collisions, the Pauli blocking is applied stochastically, with the likelihood determined by the product of the final-state blocking factors. A precise definition of these factors allows the model to account for surface effects and the depletion of the Fermi sphere as the cascade evolves. Another essential feature of INCL is its self-consistent determination of the cascade stopping time~\cite{Cugnon1997}, which is parametrized as $t_{\text{stop}} = 29.8 A_T^{0.16} \, \text{fm/c}$, where \(A_T\) is the mass number of the target nucleus. At \(t = t_{\text{stop}}\), various physical quantities—such as the excitation energy, angular momentum, and the average kinetic energy of the ejectiles—undergo a transition from a rapid evolution, driven by the intranuclear cascade, to a much slower evolution, indicative of equilibration. By adopting this self-consistent definition of stopping time, INCL avoids the need for an intermediate pre-equilibrium phase to describe the transition between the fast cascade and the evaporation-fission decay stage. Finally, cluster emission is also possible via a dynamical phase-space coalescence algorithm~\cite{ABPRC87} that permits the emission of light clusters with masses $A\leq8$.

Therefore, INCL allows for predicting the formation of spallation remnants or compound nucleus after thermalization, with their properties characterized by atomic ($Z$) and mass ($A$) numbers, strangeness number, excitation energy, and angular momentum, as shown in the Fig.~\ref{fig1} for the reaction $^{238}$U + p at 500~MeV/nucleon. Here we can see a clear correlation between the mass and charge of the compound nuclei, whereas the excitation energy and angular momentum exhibit the expected anti-correlation as both observables increase with the number of particles knocked out during the spallation reaction. 

It must be emphasized that the good agreement of INCL calculations with experimental neutron and proton emission production cross sections obtained from spallation-induced reactions on light, medium-mass, and heavy nuclei at energies from 100~MeV/nucleon up to 2~GeV/nucleon~\cite{Mancusi14,ABPRC87}, ensures accurate predictions of the excitation energy gained by the compound nucleus. The satisfactory description of isotopic cross sections~\cite{Mancusi14,JLRSPRC105,ABPRC87}, further supports the reliability of these predictions.

\subsection{ABLA++ de-excitation model}
The de-excitation model ABLA++ is employed to calculate the particle emission probabilities, which are determined using the Wei$\beta$kopf-Ewing formalism~\cite{VFWPR57}. In this framework, the decay width of a specific initial nucleus, characterized by its excitation energy ($E_i$) and angular momentum ($J_i$), into a daughter nucleus with excitation energy ($E_f$) and angular momentum ($J_f$) via the emission of a particle $\nu$ with kinetic energy $\epsilon_{\nu}$, is given by the following expression:
\begin{eqnarray}
&&\Gamma_{\nu}(E_{i}) = \frac{ (2 \cdot \theta_{\nu}+1) \cdot m_{\nu} }{\pi^{2} \cdot \hbar^{2} \cdot \rho_{i}(E_{i},J_{i}) } \times  \nonumber \\
&&\int_{0}^{E_{i}-S_{\nu}-B_{\nu}} \sigma_{c}(\epsilon_{\nu}) \cdot \rho_{f}(E_{f},J_{f}) \cdot (\epsilon_{\nu} - S_{\nu} - B_{\nu})dE_{f},\nonumber
\end{eqnarray}
where the term $\theta_{\nu}$ represents the spin of the emitted particle, $\rho_i$ and $\rho_f$ are the level densities in the initial and the daughter nucleus, respectively, $\sigma_c$ is the cross section for the inverse process, $S_\nu$ is the separation energy of the particle, $B_\nu$ is the Coulomb energy or emission barrier (only for charged particles), and $m_\nu$ the mass of the emitted particle. For a more realistic description, the separation energies and the emission barriers for charged particles are considered according to the atomic mass evaluation AME2020~\cite{mass2020} and the prescription given by Qu and collaborators~\cite{Qu2011}, respectively.

Over the past few decades, several advanced models have been developed for calculating nuclear level densities. These models employ a variety of techniques, including microscopic combinatorial methods~\cite{Hilarei06}, Hartree-Fock approaches~\cite{Nerlo06} and phenomenological analytical expressions~\cite{Krusche86}. Ideally, a microscopic approach is preferred for modeling nuclear density of states, as it provides detailed information about nuclear levels. However, the computational time required for microscopic calculations limits the practicality of this approach. Conversely, most studies related to nuclear reaction calculations favor analytical level density descriptions, as they can effectively describe experimental data for hundreds of different isotopes. In ABLA++, two phenomenological models are used for level density calculations: the constant temperature model of Gilbert-Cameron~\cite{Gilbert65} and the Fermi gas model~\cite{Bethe37}, based on the Bethe formula. In these models, the excitation energy can be adjusted to incorporate shell and pairing corrections, as outlined in Refs.~\cite{Ignat75_450,Moller95}.

Following the Fermi gas model~\cite{Bethe37}, the level density can be calculated as a function of the excitation energy $E^{*}$ and the angular momentum $J$ as:
\begin{equation}\label{eqLev}
\rho(E^{*},J) = \frac{J+1/2}{\sqrt{2\pi} \sigma^3} e^{- \frac{J(J+1)}{2\sigma^2} } \frac{\sqrt \pi}{12} \frac{e^\texttt{S}}{\widetilde{a}^{1/4}  {E^{*}}^{5/4}},
\end{equation}
where $\sigma^2$ is the spin cut-off factor given by $\sigma^2=\frac{\Im T}{\hbar^2}$ with $\Im$ as the moment of inertia of the nucleus and $T$ the nuclear temperature, $E^{*}$ is the excitation energy of the system, $\texttt{S}$ is the entropy and $\widetilde{a}$ is the asymptotic level-density parameter in units of MeV$^{-1}$. At present, the most abundant information on level densities comes from the counting of low-lying levels and from neutron resonances~\cite{Huizenga72,TvE88}. These techniques have also been extensively exploited to obtain the asymptotic level-density parameter ($\widetilde{a}$), which relates the nuclear temperature ($T$) with the excitation energy ($E^{*}$) according to $E^{*} = \widetilde{a} T^2$, and to investigate the evolution of $\widetilde{a}$ with the excitation energy~\cite{Ignatyuk21_75}. Generally, this last parameter can be written as~\cite{toke81,Ignat75}:
\begin{equation}
\widetilde{a}= \alpha_{v}  A + \alpha_{s}  B_{s} \cdot A^{2/3}+ \alpha_{k}  B_{k}  A^{1/3},\nonumber
\end{equation}
where $A$ is the mass of the nucleus and $\alpha_{v}$, $\alpha_{s}$ and $\alpha_{k}$ are the coefficients that correspond to the volume, surface and curvature components of the single-particle level densities, respectively. The values of these coefficients were calculated by Ignatyuk~\cite{Ignat75} ($\alpha_{v}$=0.095, $\alpha_{s}$=0.073, and $\alpha_{k}$=0 in units of MeV$^{-1}$) and are the most-frequently used in model calculations. In the equation, $B_{s}$ represents the ratio between the surface of the deformed nucleus and a spherical nucleus, while $B_{k}$ corresponds to the ratio between the integrated curvature of the deformed nucleus and a spherical nucleus. Their parametrizations are taken from Ref.~\cite{Myers74}.

The entropy $\texttt{S}$ is obtained according to the equation:
\begin{equation}\label{eqS}
\texttt{S} = 2 \sqrt{ \widetilde{a} E^{*}_{mod}} =  2 \sqrt{ \widetilde{a} (E^{*} + \Delta U k(E^{*}) + \Delta P h(E^{*}) ) },
\end{equation}
where $\Delta U$ is the shell-correction energy, which is calculated according to Ref.~\cite{Moller95} for excited nuclear systems. At the fission saddle point, the shell-correction energy is assumed to be negligible~\cite{Myers1996,Karpov2008}. The term $k(E^{*})$ describes the damping of the shell effect with the excitation energy, and is calculated according to Ref.~\cite{Ignatyuk21_75} as:
\begin{equation}
k(E^{*}) = 1 - e^{-\gamma E^{*}}, \nonumber
\end{equation}
with the parameter $\gamma$ given by $\gamma = \widetilde{a}/(0.4A^{4/3})$ in units of MeV$^{-1}$~\cite{Khs1982}. The term $\Delta P$, which is identical to the pairing condensation energy in odd-odd nuclei, is calculated as:
\begin{equation}
\Delta P = 2\delta + 1/4 g \delta^2,\nonumber
\end{equation}
with an average pairing gap of $\delta = 12/\sqrt{A}$ in units of MeV, and with a single-particle level density at the Fermi energy of $g = 6 \widetilde{a}/\pi^2$. Finally, the term $h(E^{*})$ describes the superfluid phase transition~\cite{Ignat75_450} according to Ref.~\cite{Ignatyuk1977} as:
\begin{equation}
h(E^{*})=
\begin{cases}
  1 - (1-\frac{E^{*}}{E_{crit}})^2 & \text{if }E^{*}<E_{crit}\\
  1 & \text{if }E^{*}\geq E_{crit}
\end{cases}\text, \nonumber
\end{equation}
where the critical energy ($E_{crit}$) is set to 10 MeV. Note that in these equations the excitation energy is shifted according to the prescription given in Ref.~\cite{Abla07} to accommodate for the different energies of even-even, odd-mass, and odd-odd nuclei.

To calculate the intrinsic level density at very low excitation energies, ABLA++ switches from the Fermi-gas level density to the constant-temperature level density~\cite{Gilbert65}. The calculation is based on the work performed in Ref.~\cite{Ignatyuk2001}, where the values of the parameters of the constant-temperature level density approach were determined through the simultaneous analysis of neutron resonances and low-lying levels within the framework of the Gilbert-Cameron model~\cite{Gilbert65}.

To account for the role of collective excitations in the decay of excited compound nuclei, the level density of Eq.~\ref{eqLev} is corrected using the vibrational and rotational enhancement factors according to:
\begin{equation}\label{eqLevColl}
\rho(E,J) =  K_{vib} K_{rot} \rho(E,J)_{int},\nonumber
\end{equation}
where $\rho(E,J)_{int}$ is given by Eq.~\ref{eqLev}, $K_{vib}$ represents the vibrational enhancement factor and $K_{rot}$ corresponds to the rotational factor. For nuclei with highly deformed saddle point or significant ground-state deformation, the collective enhancement factor arises from the formation of rotational bands above the intrinsic single-particle levels. In this case, the vibrational factor $K_{vib}$ can be considered negligible, while the rotational enhancement factor is calculated according to Refs.\cite{Bjornholm,Hui74a} in terms of the rigid-body moment of inertia. In contrast, for spherical nuclei, collective motion is described based on low-frequency vibrational modes. Currently, the factor $K_{vib}$ can be computed from the statistical sum of harmonic vibrational modes~\cite{Bjornholm} or using phenomenological approaches~\cite{Junghans98}. In ABLA++, these factors are calculated based on the phenomenological description proposed by Junghans and collaborators~\cite{Junghans98}.

For nuclei with a quadrupole deformation $|\beta_2| > 0.15$, the rotational enhancement factor $K_{rot}(E^{*}_{mod})$ is calculated in terms of the spin-cutoff parameter $\sigma_\bot$:
\begin{equation}\label{eqrot}
K_{rot}(E^{*}_{mod}) =
\begin{cases}
  1+(\sigma_\bot^2-1)f(E^{*}_{mod}) & \text{if } \sigma_\bot > 1\\
  1 & \text{if } \sigma_\bot \leq 1
\end{cases}\text,
\end{equation}
being $\sigma_\bot^2 = \frac{\Im_\bot T}{\hbar^2}$, where $T$ is the temperature of the nuclear system and $\Im_\bot$ is the rigid-body moment of inertia perpendicular to the symmetry axis determined according to Ref.~\cite{Hasse}. In this expression $E^{*}_{mod}$ is defined by Eq.~\ref{eqS}. The ground-state quadrupole deformation $\beta_2$ is taken from the finite-range liquid-drop model including microscopic corrections~\cite{Moller95}, while the saddle-point deformation is taken from the liquid-drop model as given in Ref.~\cite{Cohen1963}. The damping of the collective modes with increasing excitation energy ($f(E^{*}_{mod})$) is described by a Fermi function as:
\begin{equation}
f(E^{*}_{mod}) = \left(1 + e^{\frac{E^{*}_{mod}-E_c}{d_c}}\right)^{-1} ,\nonumber
\end{equation}
with the parameters $E_c$ = 40 MeV and $d_c$ = 10 MeV. The vibrational enhancement for spherical nuclei is generally smaller than the rotational enhancement for deformed nuclei. For nuclei with a quadrupole
deformation $|\beta_2| \leq 0.15$, the vibrational enhancement factor is calculated by using the same formula as for the rotational enhancement (Eq.~\ref{eqrot}), but with the spin-cutoff parameter calculated assuming an irrotational flow:
\begin{equation}
\sigma_\bot'^{2} = 75\beta_{eff} \sigma_\bot^{2},\nonumber
\end{equation}
where $\beta_{eff}$ is a dynamical deformation parameter expressed as $\beta_{eff} = 0.022 + 0.003\Delta N + 0.002\Delta Z$, being $\Delta N$ and $\Delta Z$ the absolute values of the number of neutrons and protons, respectively, above or below the nearest shell closure.

As is well known in the standard Wei$\beta$kopf-Ewing approach~\cite{VFWPR57}, the change of angular momentum in the evaporation process due to particle emission is not addressed. To overcome this limitation, a dedicated formalism, as described in Ref.~\cite{Abla07}, is incorporated into ABLA++ to calculate the distribution of orbital angular momentum during the emission of particles from excited nuclear systems with finite angular momentum.

Furthermore, the inverse cross section ($\sigma_{c}$) for the emission of particles is calculated considering several effects: (i)~the existence of the Coulomb barrier for charged particles (especially at low energy), (ii)~the tunnelling through it (especially for light particles), and (iii)~the energy-dependent quantum-mechanical cross section. At energies well above the Coulomb barrier, where the shape of the barrier no longer plays a role, $\sigma_{c}$ can then be calculated without considering tunnelling as~\cite{Abla07}:
\begin{equation}
\sigma_{c}(\epsilon_{\nu}) = \pi \left(1.16 (A_1^{1/3}+A_2^{1/3}) + \sqrt{ \frac{\hbar^2}{2\mu E_{cm}} }\right)^2 \left( 1 - \frac{B_{\nu}}{\epsilon_{\nu}} \right),\nonumber
\end{equation}
where $\mu$ is the reduced mass, calculated as $\mu = M_1 M_2 /(M_1+M_2)$, and $E_{cm} = \epsilon_{\nu} (A_1 - A_2)/A_1$.
\begin{figure}[t!]
\centering
\subfigure{\label{fig2a}\includegraphics[width=0.486\textwidth,keepaspectratio]{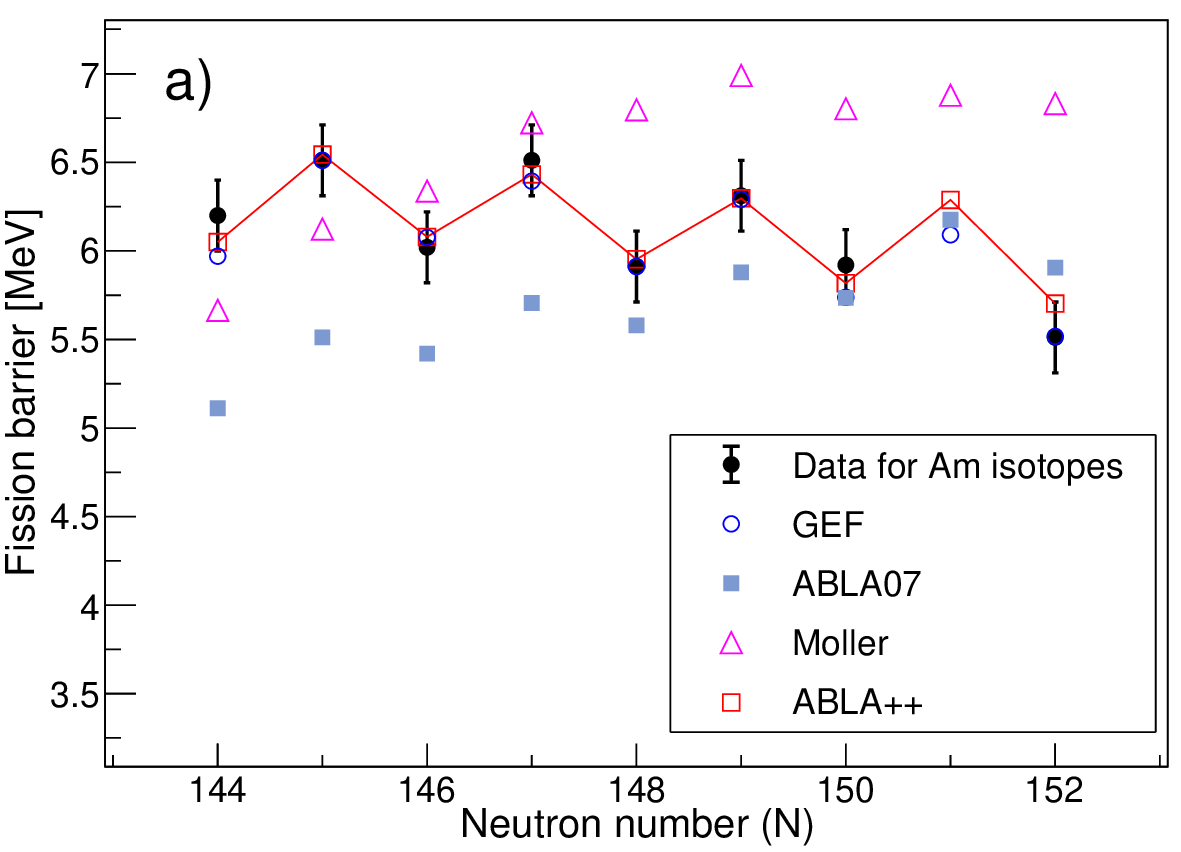}}
\centering
\subfigure{\label{fig2a}\includegraphics[width=0.486\textwidth,keepaspectratio]{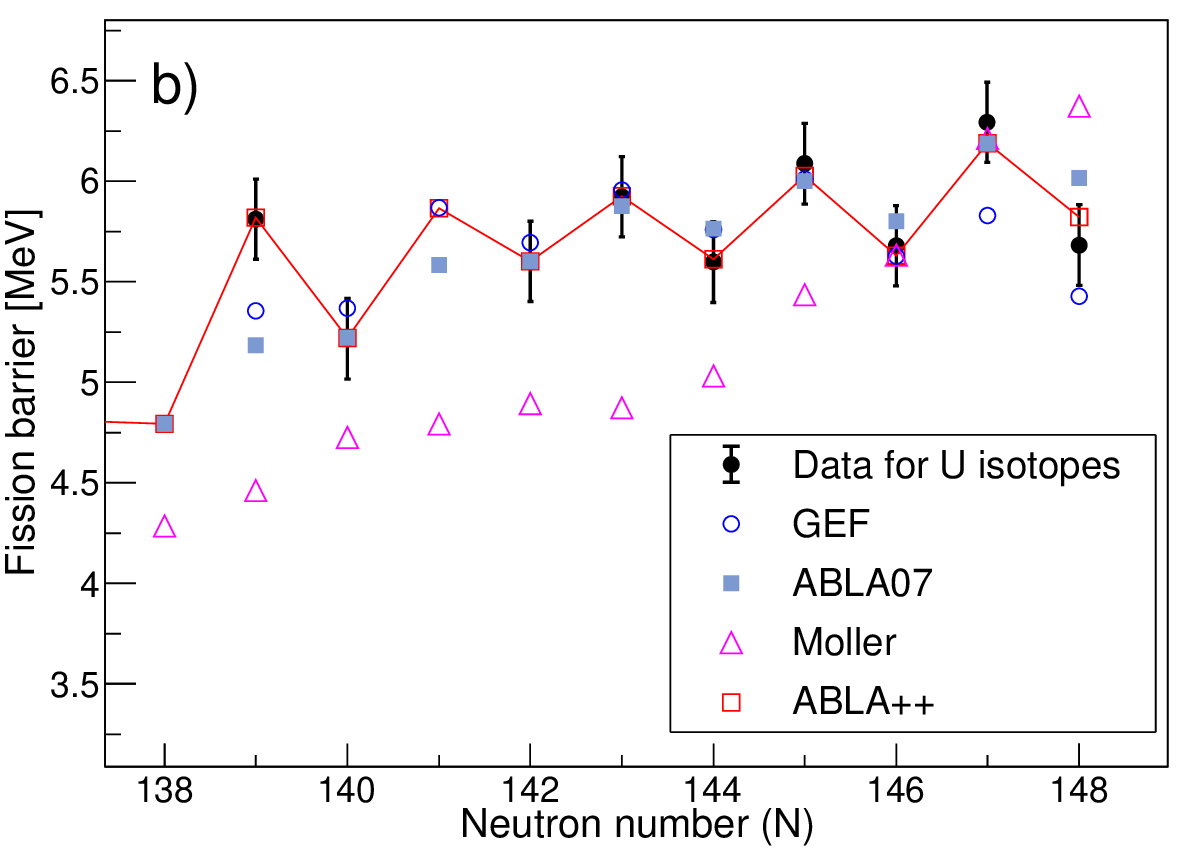}}
\centering
\subfigure{\label{fig2b}\includegraphics[width=0.486\textwidth,keepaspectratio]{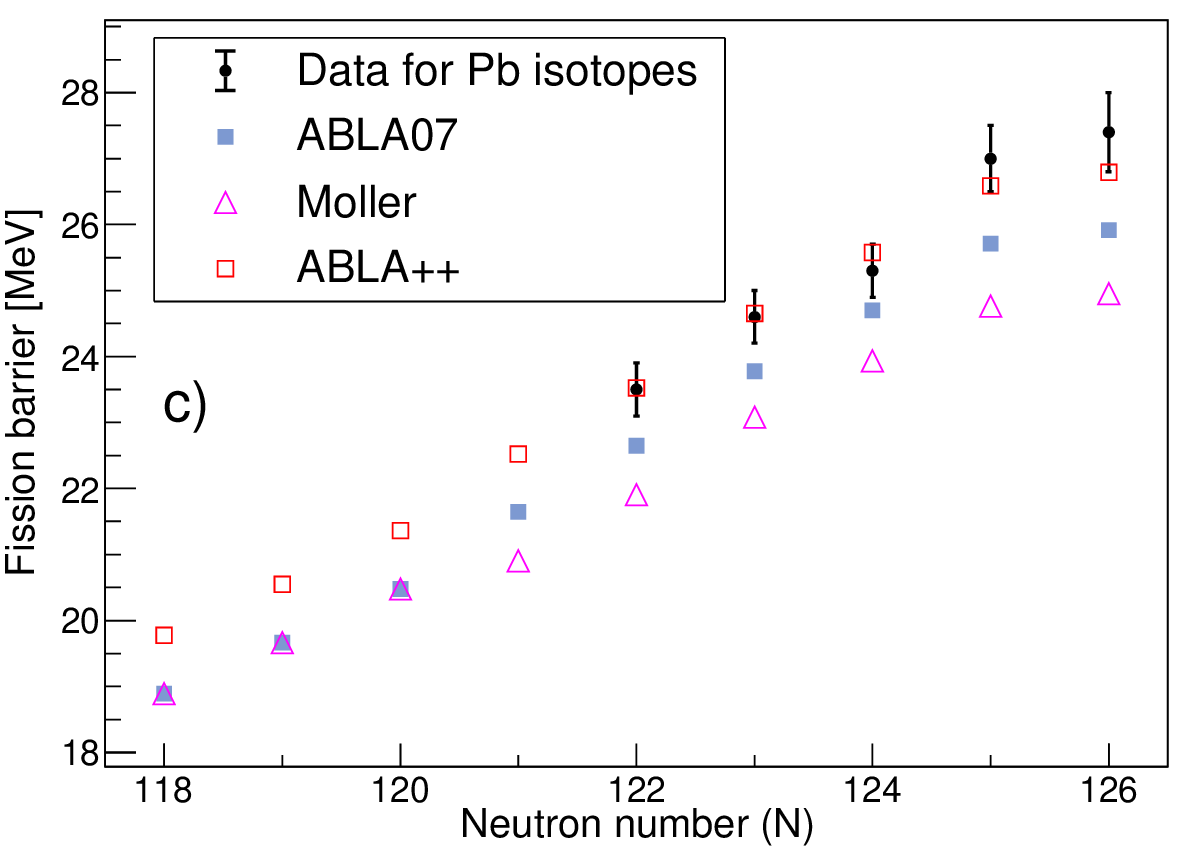}}
\caption{(Color online) Fission-barrier heights above the nuclear ground state for americium (a), uranium (b), and lead (c) isotopes. The ABLA++ model predictions (open squares) are compared with empirical fission barriers (filled circles), as well as to results from the GEF~\cite{GEF16} model (open circles), ABLA07~\cite{Abla07} model (filled squares), and the theoretical calculations by M\"{o}ller and collaborators~\cite{Moller2015} (open triangles).
}
\label{fig2}
\end{figure}

Given the significant role of fission in the decay of heavy nuclei, at each de-excitation stage, a competition between fission and other particle decay channels is evaluated. In ABLA++, the fission decay width is described by the Bohr-Wheeler transition-state model~\cite{NBPR56}, following the formulation provided by Moretto~\cite{LGMNPA247}:
\begin{equation}\label{eqbw}
\Gamma_{f}^{BW}= \frac{T}{2  \pi}  \frac{\rho_{sp}(E-B_f,J)}{\rho_{gs}(E,J)},
\end{equation}
where $\rho_{sp}(E-B_f,J)$ and $\rho_{gs}(E,J)$ represent the level densities at the saddle-point and ground-state configurations, respectively, $T$ is the nuclear temperature, and $B_f$ is the fission-barrier height obtained from the finite-range liquid-drop model of Sierk~\cite{AJSPRC33}, accounting for the influence of angular momentum and incorporating the ground-state shell effects~\cite{Moller95}. In Fig.~\ref{fig2}, a comparison of the fission-barrier heights used by the GEF~\cite{GEF16}, ABLA07, and the newer ABLA++ model is displayed for americium, uranium, and lead isotopes, alongside experimental data~\cite{Bjornholm80,Dahlinger1982} and the predictions from M\"{o}ller's model~\cite{Moller2015}. On the one hand, the theoretical values from ABLA07~\cite{Abla07} are in close agreement with the experimental data, whereas the values provided by M\"{o}ller and collaborators show significant deviations both in their absolute values and in the isotopic trends. On the other hand, the parametrization from GEF and the current one implemented in ABLA++ provide very good description of the empirical data.
\begin{figure}[t!]
\centering
\subfigure{\label{fig3a}\includegraphics[width=0.49\textwidth,keepaspectratio]{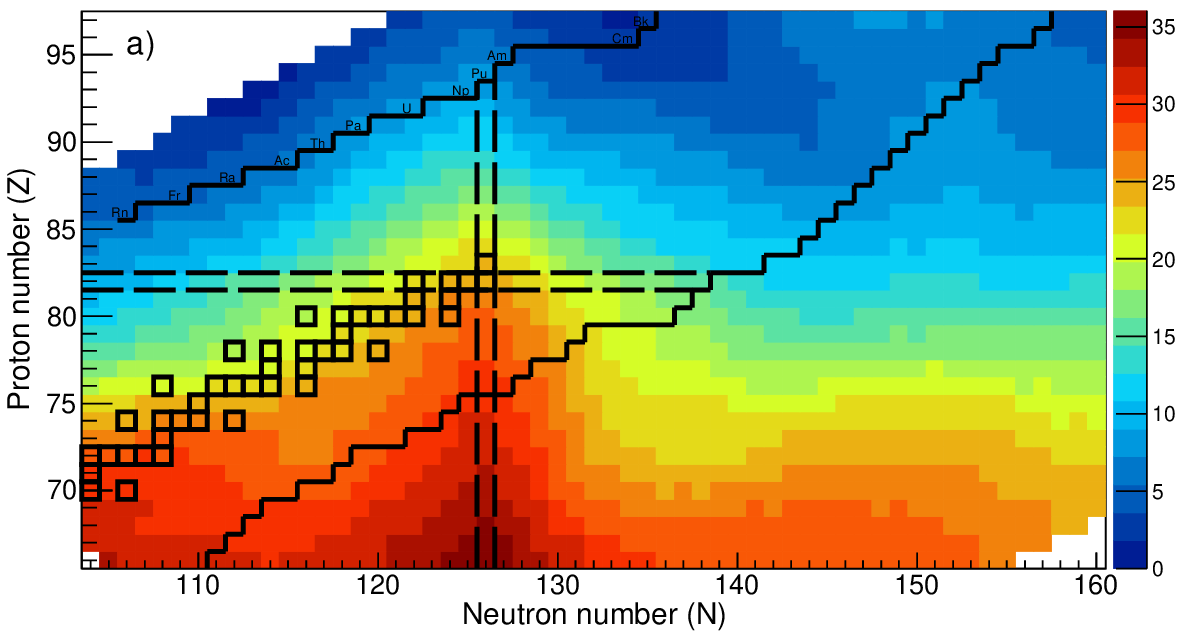}}
\centering
\subfigure{\label{fig3b}\includegraphics[width=0.49\textwidth,keepaspectratio]{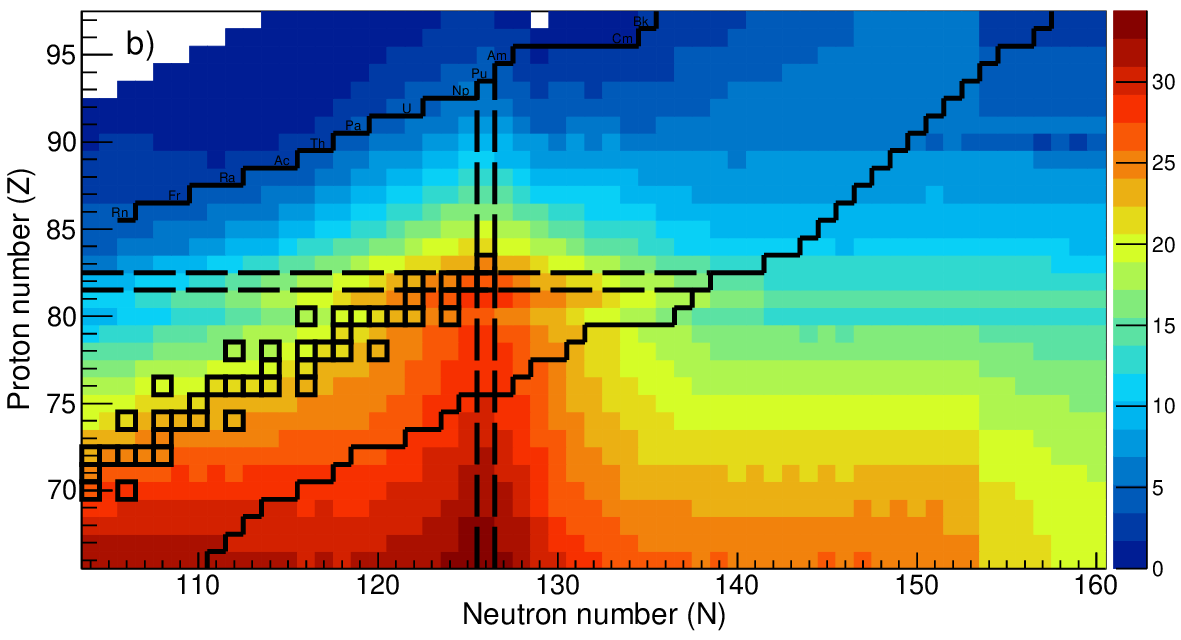}}
\caption{(Color online) Chart of fission-barrier heights (in MeV) as a function of the fissioning system in terms of proton and neutron numbers. (a) Macroscopic-microscopic finite-range liquid-drop
model predictions developed by M\"{o}ller and collaborators~\cite{Moller2015} and (b) results from the modified finite-range liquid-drop model of Sierk~\cite{AJSPRC33} implemented in ABLA++. For orientation, the primordial stable isotopes are indicated by black open squares. The limits of known nuclei were obtained from the atomic mass evaluation AME2020~\cite{mass2020}.
}
\label{fig3}
\end{figure}
Finally, in Fig.\ref{fig3}, the fission-barrier heights for all the heavy nuclei are displayed, comparing the results from M\"{o}ller et al. (Fig.\ref{fig3a}) with the ABLA++ parametrization (Fig.~\ref{fig3b}). Both models exhibit a very similar mapping of the fission barriers, with a clear maximum for nuclei around $N=126$.

The dynamical diffusion process above the fission barrier is described by the Fokker-Planck equation, for which the quasi-stationary solution was introduced by Kramers~\cite{kramers1940}, leading to a reduction of the fission decay width due to dissipation:
\begin{equation}
\Gamma_{f}^{K} = \left[ \sqrt{1+ \left(\frac{\beta}{2 \omega_{0}}\right)^2} -\frac{\beta}{2 \omega_{0}} \right] \Gamma_{f}^{BW}.\nonumber
\end{equation}
Here $\beta$ is the dissipation coefficient and $\omega_{0}$ is the frequency of the harmonic oscillator that describes the inverted potential at the fission barrier, calculated using the liquid-drop model~\cite{Nix1967}. This equation provides the asymptotic value of the fission decay width, which is then modified through an analytical approximation to the solution of the one-dimensional Fokker-Planck equation~\cite{FPE_book}, as developed by Jurado and collaborators in Refs.\cite{BJPLB553,BJNPA747}, to account for the time dependence of the fission-decay width. This description was later refined to also include the initial quadrupole deformation of the compound nucleus, yielding a more realistic representation of the fission process in the actinide region\cite{schmitt_schmidt}. Under this approximation, the time-dependent fission-decay width is defined as:
\begin{equation}
\Gamma_f(t)= \frac{W_n (x = x_{sd} ; t, \beta)}{W_n (x = x_{sd} ; t\rightarrow \infty, \beta)} \Gamma_{f}^{K},\nonumber
\end{equation}
where $W$($x; t, \beta$) is the normalized probability distribution at the saddle-point deformation $x_{sd}$, being the saddle-point deformations calculated according to Ref.~\cite{Hasse}.

In the case of a nuclear potential approximated by a parabolic shape, the solution to the Fokker-Planck equation for the probability distribution $W$($x; t, \beta$) at the saddle-point deformation takes a Gaussian form, with a time-dependent width~\cite{Chandrasekhar}. The zero-point motion is incorporated by shifting the time scale by a certain amount $t_0$, which represents the time required for the probability distribution to achieve the width corresponding to the zero-point motion in deformation space. This time is computed as~\cite{BJPLB553}:
\begin{equation}
 t_0 =
\begin{cases}
 \frac{1}{\beta} ln\left( \frac{2T}{2T-\hbar \omega_1} \right) & \text{if } \beta \leq 2\omega_1 \\
 \frac{\hbar  \beta}{4\omega_1 T} & \text{if } \beta > 2\omega_1
\end{cases}\text,\nonumber
\end{equation}
where $\omega_1$ describes the curvature of the potential at the ground state.
\begin{figure}[b!]
\centering
\subfigure{\includegraphics[width=0.49\textwidth,keepaspectratio]{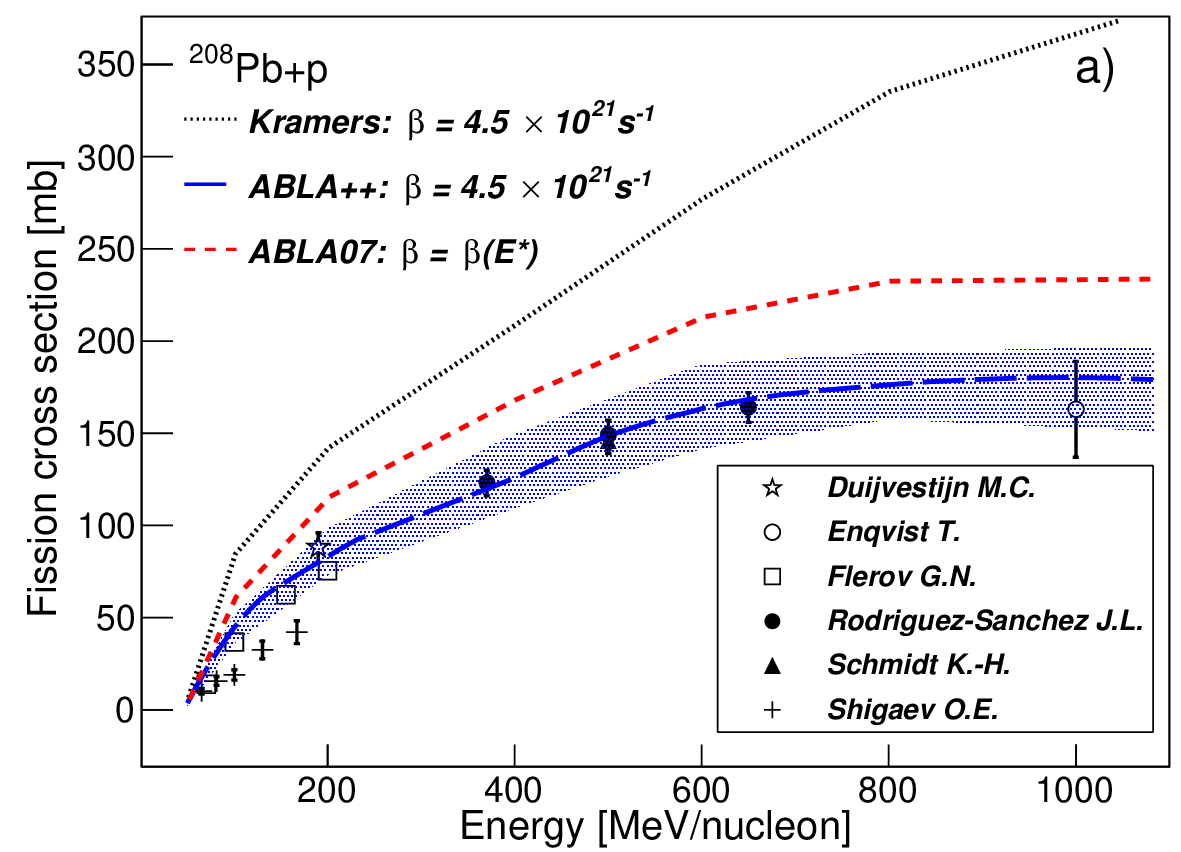}\label{fig4a}}
\centering
\subfigure{\includegraphics[width=0.49\textwidth,keepaspectratio]{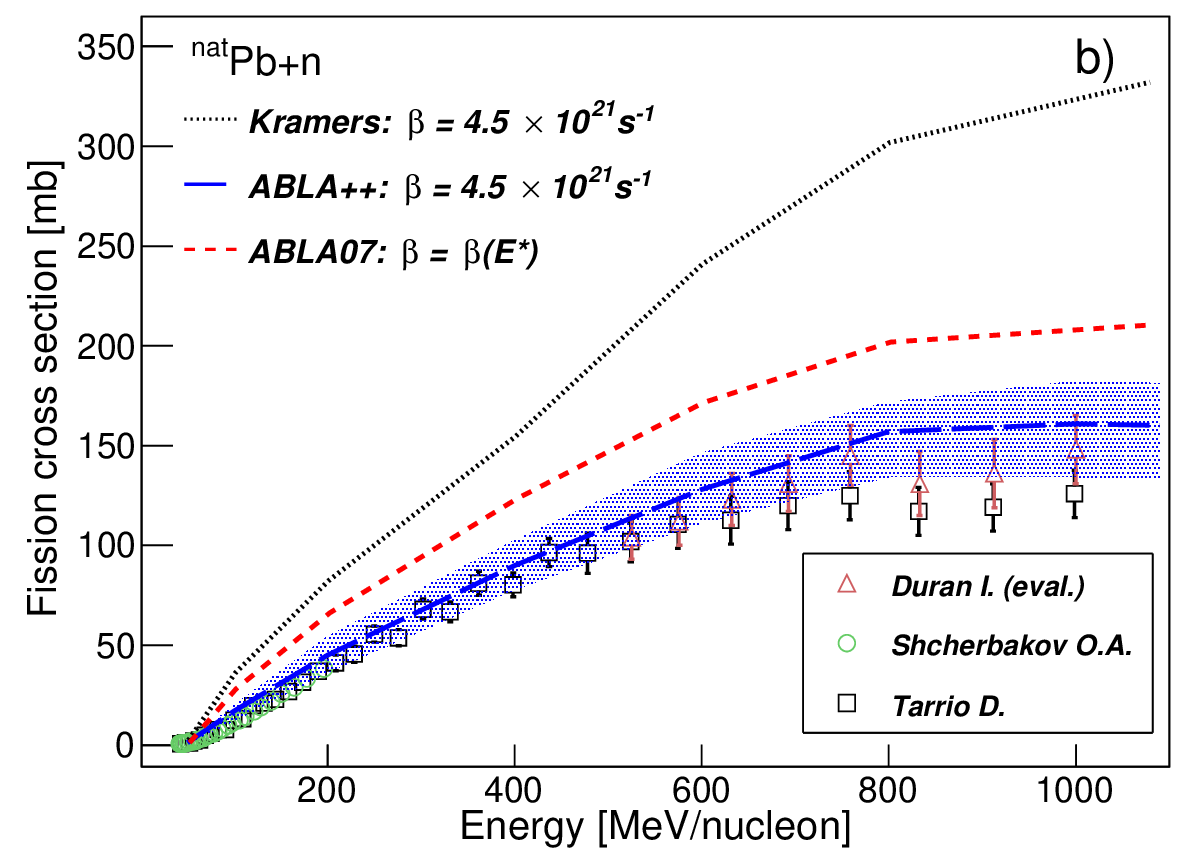}\label{fig4b}}
\caption{(Color online) Proton-induced total fission cross sections measured in different works as a function of the incident kinetic energy for the reactions $^{208}\mathrm{Pb}$ + p~\cite{EXFORWEB,Schmidt13,JLRS14} (a) and $^{\text{nat}}\mathrm{Pb}$ + n~\cite{EXFORWEB,Tarrio2011,IDEPJW146} (b). The lines correspond to Kramers' approach (dotted line), ABLA++ (long-dashed line), and ABLA07 (short-dashed line). See text for more details.
}
\label{fig4}
\end{figure}

\afterpage{%
\begin{figure*}[t]
\centering
\subfigure{\includegraphics[width=0.97\textwidth,keepaspectratio]{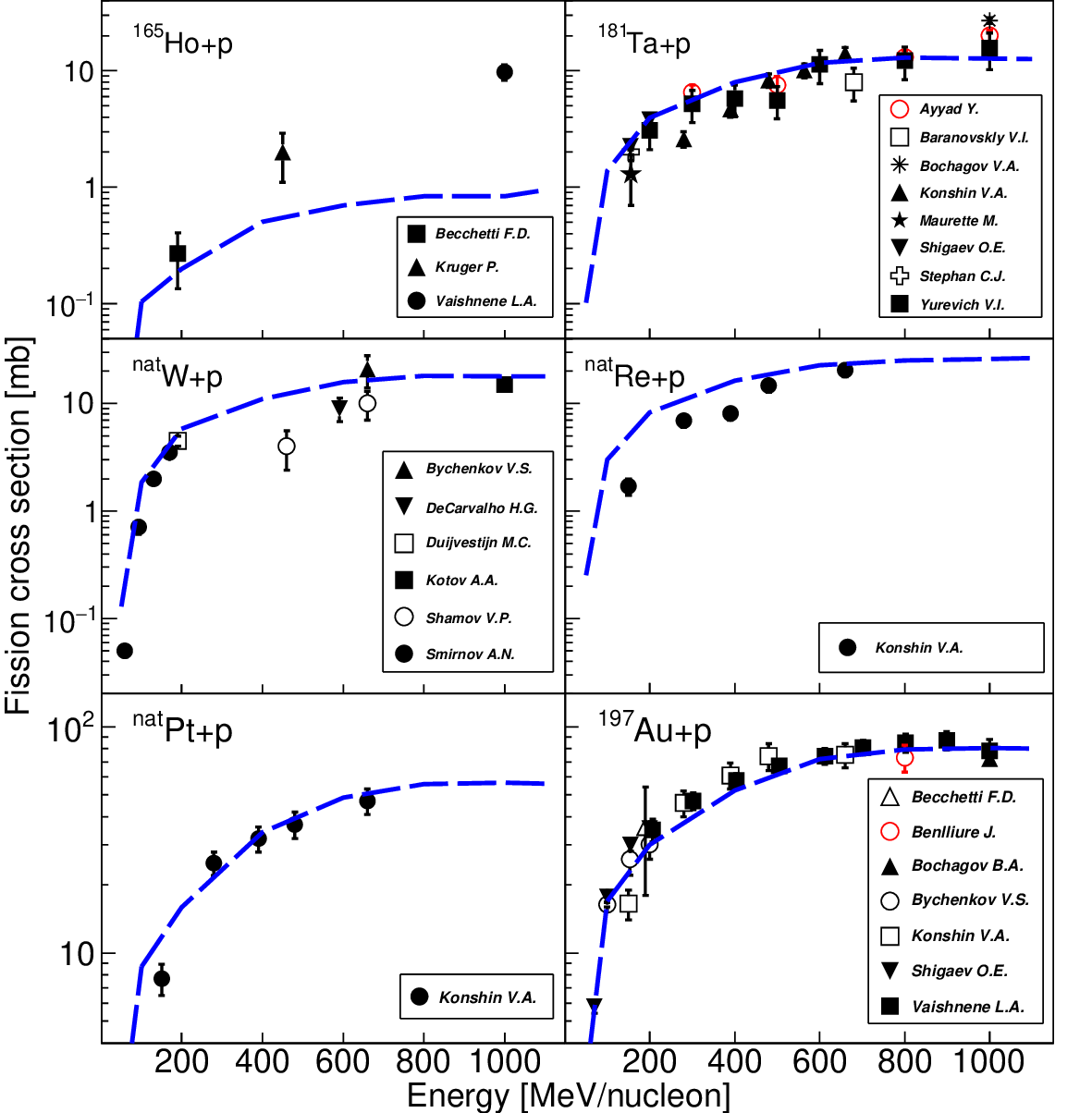}}
\caption{(Color online) Proton-induced total fission cross sections measured in direct (black symbols) and inverse (colored symbols) kinematics for $^{165}\mathrm{Ho}$, $^{181}\mathrm{Ta}$, $^{\text{nat}}\mathrm{W}$, $^{\text{nat}}\mathrm{Re}$, $^{\text{nat}}\mathrm{Pt}$, and $^{197}\mathrm{Au}$ nuclei as a function of the incident energy~\cite{EXFORWEB,Yassid89_2014}. The dashed lines represent calculations performed using the INCL-ABLA++ models.}
\label{fig5}
\end{figure*}

\begin{figure*}[t]
\centering
\subfigure{\includegraphics[width=0.97\textwidth,keepaspectratio]{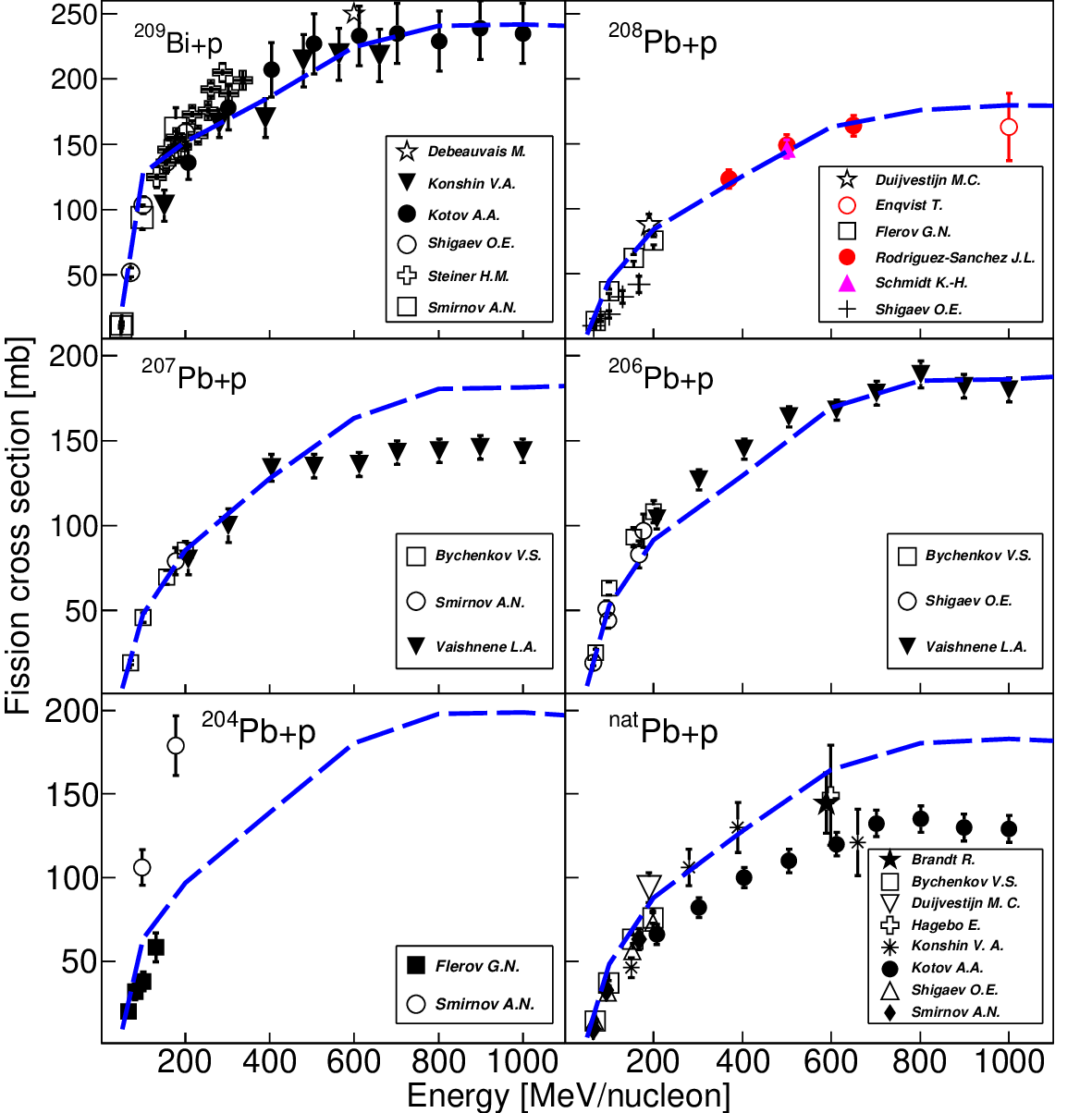}}
\caption{(Color online) As in Fig.~\ref{fig5}, but for the $^{209}\mathrm{Bi}$, $^{208,207,206,204}\mathrm{Pb}$, and $^{\text{nat}}\mathrm{Pb}$ nuclei~\cite{EXFORWEB,Schmidt13,YAPRC91,JLRS14}.
}
\label{fig6}
\end{figure*}

\begin{figure*}[t]
\centering
\subfigure{\includegraphics[width=0.97\textwidth,keepaspectratio]{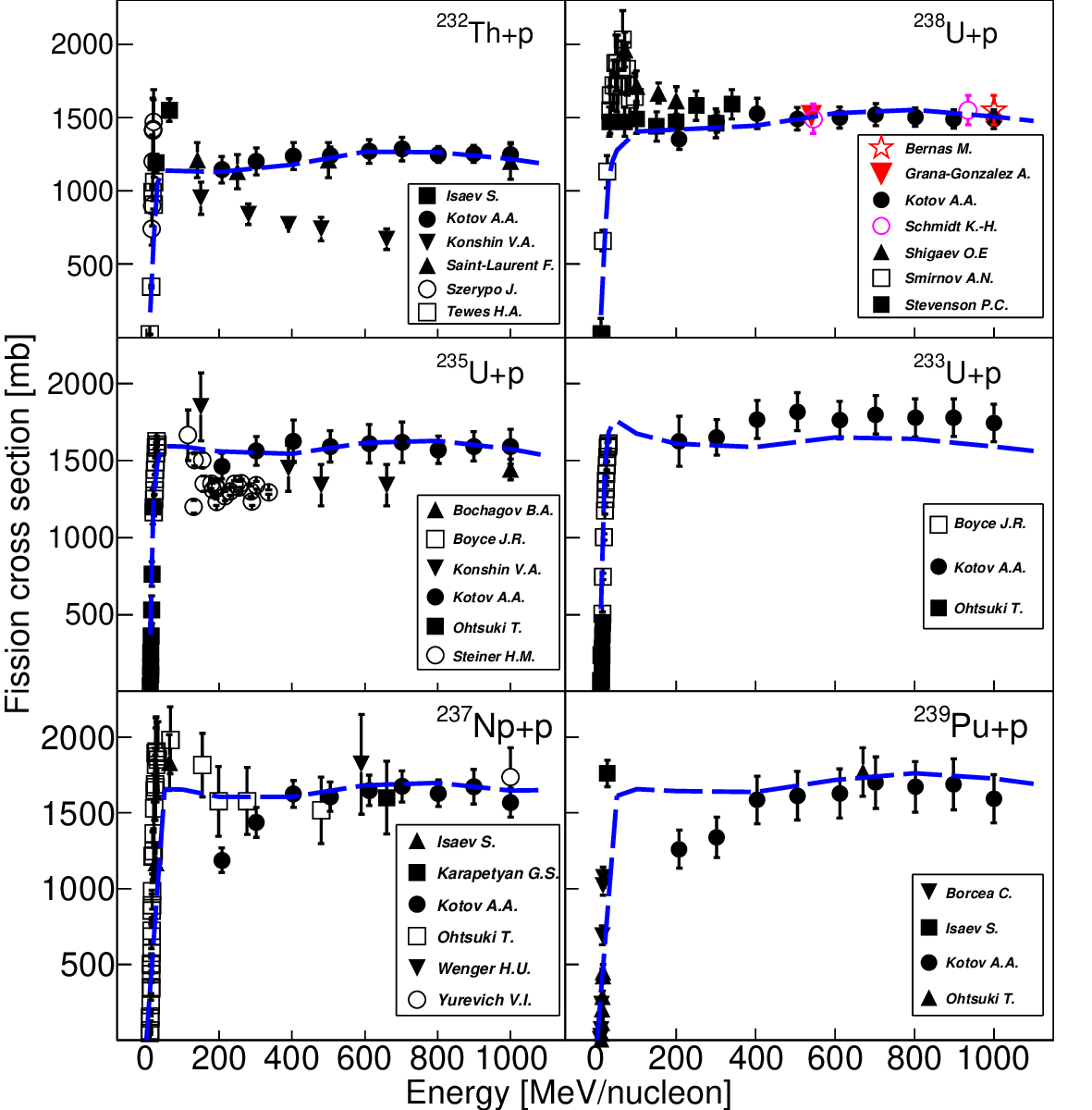}}
\caption{(Color online) As in Fig.~\ref{fig5}, but for the $^{232}\mathrm{Th}$, $^{238,235,233}\mathrm{U}$, $^{237}\mathrm{Np}$, and $^{239}\mathrm{Pu}$ nuclei~\cite{EXFORWEB,Bernas2003,Kotov2006,Schmidt13,Antia_thesis}.
}
\label{fig7}
\end{figure*}

\begin{figure*}[t]
\centering
\subfigure{\includegraphics[width=0.97\textwidth,keepaspectratio]{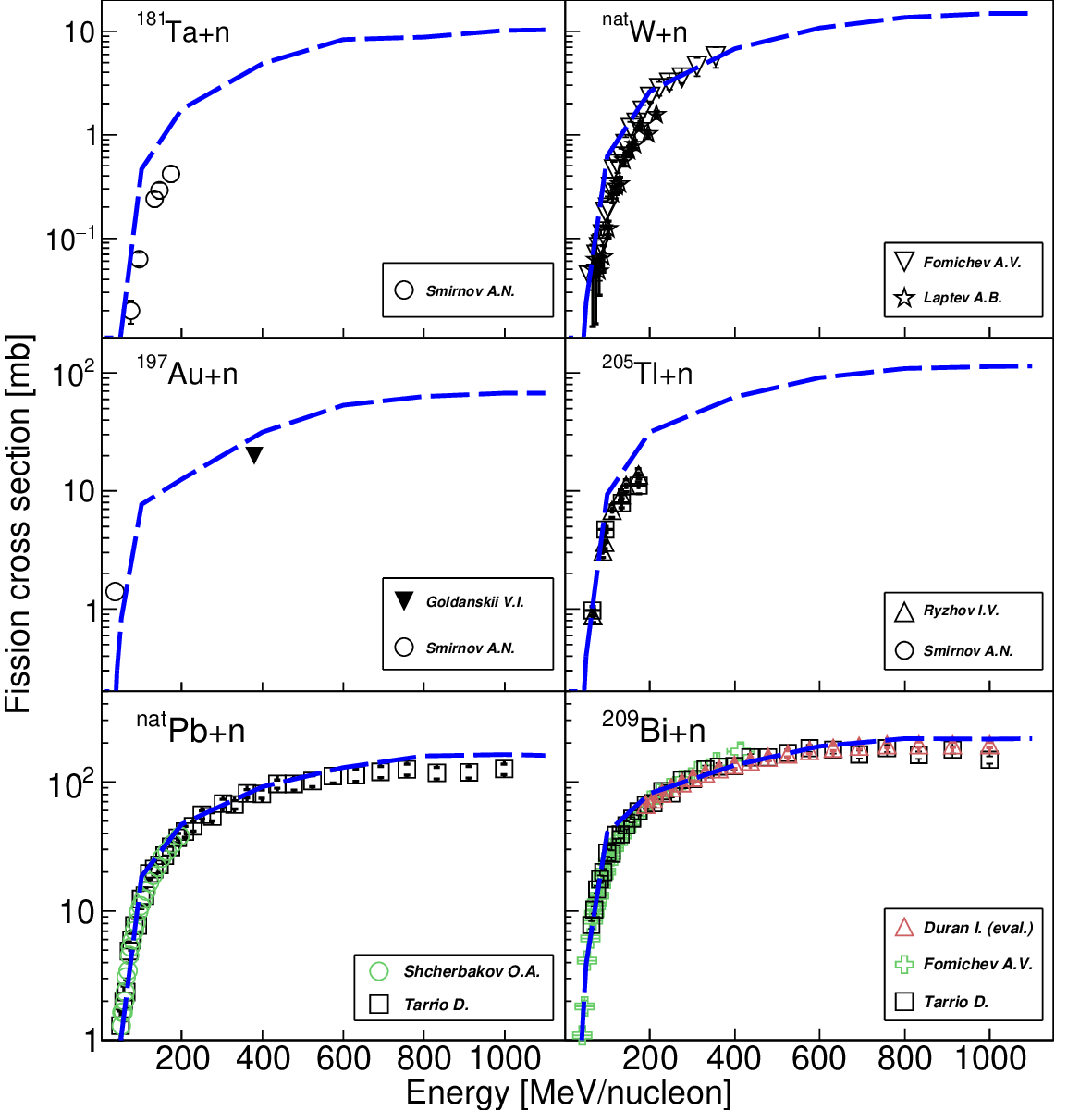}}
\caption{(Color online) Neutron-induced total fission cross sections measured in direct kinematics for $^{181}\mathrm{Ta}$, $^{\text{nat}}\mathrm{W}$, $^{197}\mathrm{Au}$, $^{205}\mathrm{Tl}$, $^{\text{nat}}\mathrm{Pb}$, and $^{209}\mathrm{Bi}$ nuclei as a function of the incident energy~\cite{EXFORWEB,IDEPJW146,Tarrio2011}. The dashed lines represent the INCL-ABLA++ calculations.
}
\label{fig8}
\end{figure*}

\begin{figure*}[t]
\centering
\subfigure{\includegraphics[width=0.97\textwidth,keepaspectratio]{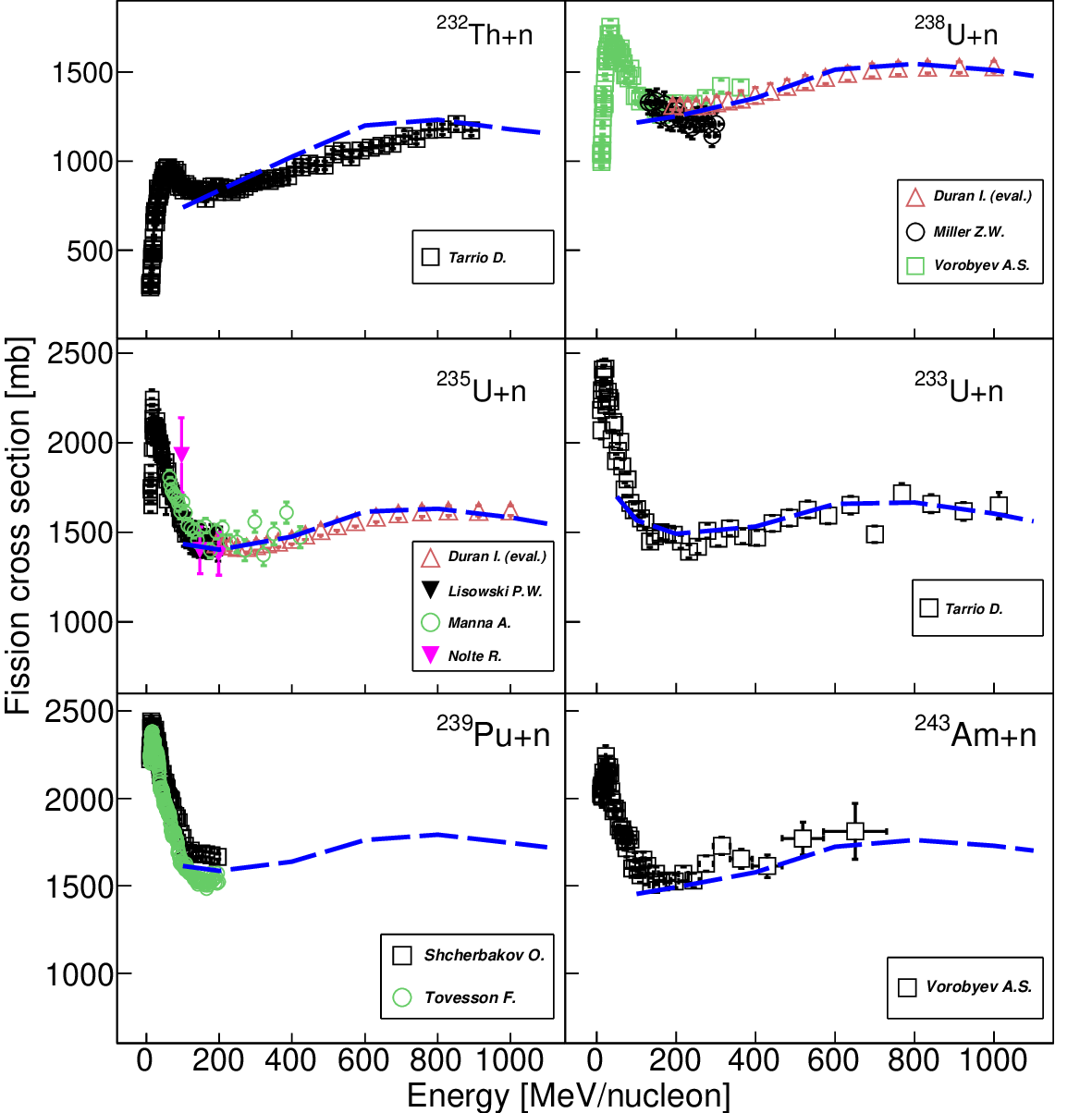}}
\caption{(Color online) As in Fig.~\ref{fig8}, but for the $^{232}\mathrm{Th}$, $^{238,235,233}\mathrm{U}$, $^{239}\mathrm{Pu}$, and $^{243}\mathrm{Am}$
 nuclei~\cite{EXFORWEB,IDEPJW146,Tarrio2023,Manna2025}.
}
\label{fig9}
\end{figure*}
}
This then leads to the following analytical approximation to the solution of the Fokker-Planck equation for the time-dependent fission width:
\begin{equation}
\Gamma_f(t)= \frac{K_BT}{\mu \omega_1^2 \sigma^2}\frac{e^{-\frac{(x_{sd}-x_{m})^2}{2\sigma^2}}}{ e^{-\frac{\mu \omega_{1}^2 x_{sd}^2}{2K_BT}} } \Gamma_{f}^{K},\nonumber
\end{equation}
with $\sigma^2$ given as:
\begin{eqnarray}
&&\sigma^2 = \frac{K_BT}{\mu \omega_1^2} [ 1 - e^{-\beta(t+t_0)} \times \nonumber \\
&&\left( \frac{2\beta^2}{\beta_1^2} sinh^2( 0.5 \beta_1 (t+t_0)) + 
\frac{\beta}{\beta_1}sinh( \beta_1(t+t_0) + 1) \right) ] ,\nonumber 
\end{eqnarray}
where $K_B$ denotes Boltzmann's constant, $\mu$ represents the reduced mass associated with the deformation degree of freedom, and $\beta_1 =(\beta^2-4\omega_1^2)^{1/2}$. The dissipation coefficient $\beta$ is set to $4.5\times10^{-21}$ s$^{-1}$, a value constrained by multiple studies~\cite{schmitt_schmidt,JLRS14,YAPRC91,JLRS15_2}. Finally, the mean deformation $x_m$ of the system is calculated as:
\begin{equation}
 x_m =
\begin{cases}
X_{i}cos[0.5\beta_2(t-t_0)]e^{-\beta(t-t_0)} & \text{if } \beta \leq 2\omega_1\\
X_{i}e^{-0.5(\beta-\beta_1)(t-t_0)} & \text{if }  \beta > 2\omega_1
\end{cases}\text,\nonumber
\end{equation}
where the initial deformation $X_{i}$ is calculated according to Ref.~\cite{Myers74}.

In low-energy fission, the double-humped structure observed in the fission barrier as a function of elongation, along with the symmetry classes at various saddle points, plays an important role in accurately describing the fission process. These effects have also been incorporated into the ABLA++ model~\cite{Abla07}, based on the theoretical frameworks developed in Refs.~\cite{Gavron76,Strutinsky68,Bjornholm80}.

In Figs.~\ref{fig4a} and~\ref{fig4b} different ABLA++ calculations are compared with experimental total fission cross sections of the reactions $^{208}\mathrm{Pb}$ + p and $^{nat}\mathrm{Pb}$ + n, respectively, displayed as a function of the incident kinetic energy. The dotted line represents Kramers' calculations considering a constant dissipation coefficient $\beta=4.5 \times 10^{-21}$ s$^{-1}$, whereas the long-dashed line corresponds to complete dynamical ABLA++ calculations, which also account for transient time effects. Additionally, the results from ABLA07 are also shown (short-dashed line), which by default include an excitation-energy-dependent dissipation coefficient parametrized by the following expression:
\begin{equation}
\beta = \beta_0 \left( 1 - e^{-2.5 \widetilde{a} E^* A^{-\frac{4}{3}}} \right) ,\nonumber
\end{equation}
where $\beta_0$ is set to the standard value of $4.5\times10^{-21}$ s$^{-1}$, $\widetilde{a}$ is the level density parameter, $E^*$ is the excitation energy above the saddle point, and $A$ corresponds to the mass number of the fissioning nucleus at the saddle point.

Kramers' calculations clearly overestimate the experimental cross sections, whereas the ABLA++ calculations including transient time effects results in a very good agreement with the empirical data. The transient time effects manifest mostly at high kinetic energies ($>$70~MeV/nucleon), where the transient time exceeds the statistical evaporation time~\cite{BJPRL93,schmitt_schmidt,Yassid89_2014}. In the case of ABLA++ calculations, the dashed area represents the fission cross-section range covered by an asymmetric uncertainty of $[-0.5, +1]$ for the dissipation coefficient $\beta$, explicitly encompassing values between 4 and $5.5 \times 10^{-21}$ s$^{-1}$, respectively. This dashed area mostly covers all the fission cross sections in the energy range between 50 and 1100~MeV/nucleon. However, the calculations considering the excitation-energy-dependent dissipation coefficient significantly overestimate the experimental data, as this parametrization reduces the dissipation parameter's value for fission reactions at moderate excitation energies, specifically between the fission-barrier height and 100~MeV. Consequently, this excitation-energy dependence was excluded in the ABLA++ model.

\section{Results}
After fixing the fission-barrier heights and dissipation coefficient in ABLA++, the INCL-ABLA++ calculations are benchmarked against existing fission cross-section datasets for neutron- and proton-induced fission reactions and incident kinetic energies ranging from 20 to 1100~MeV/nucleon. For proton-induced fission reactions, experimental data are available in both direct and inverse kinematics, whereas for neutron-induced fission reactions, data only exist in direct kinematics.

Figs.~\ref{fig5}, ~\ref{fig6}, and ~\ref{fig7} display total fission cross sections induced by protons as a function of the incident energy for light pre-actinides, pre-actinides, and actinides, respectively. Most of the data points were taken from the EXFOR database~\cite{EXFORWEB}. As can be seen in Fig.~\ref{fig5}, INCL-ABLA++ calculations provide a good description of proton-induced total fission cross sections for $^{181}\mathrm{Ta}$, $^{\text{nat}}\mathrm{W}$, $^{\text{nat}}\mathrm{Re}$, $^{\text{nat}}\mathrm{Pt}$, and $^{197}\mathrm{Au}$ nuclei as a function of incident energy. The calculations slightly underestimate the experimental data for $^{165}\mathrm{Ho}$. However, this discrepancy might arise from an overestimation of fission events as other reaction mechanisms, such as multi-fragmentation and intermediate-mass fragment (IMF) emission, may open up at high excitation energies~\cite{Natowitz2002} and generate similar coincidence signals in the fission detectors, leading to contamination of the identified fission events. Moreover, other data comparisons show that the existing data sometimes exhibit large fluctuations, particularly for cross sections around 1~mb. Therefore, additional measurements are needed to confirm the experimental trend.

Fig.~\ref{fig6} displays the experimental data for the $^{209}\mathrm{Bi}$, $^{208,207,206,204}\mathrm{Pb}$, and $^{\text{nat}}\mathrm{Pb}$ nuclei~\cite{EXFORWEB,Schmidt13,YAPRC91,JLRS14} measured in both direct and inverse kinematics. In this case, the INCL-ABLA++ model provides an excellent description for the $^{209}\mathrm{Bi}$ and $^{208}\mathrm{Pb}$ nuclei. However, for the other nuclei, the quality of the agreement depends on the specific dataset. The measurement performed by Kotov et al. for $^{\text{nat}}\mathrm{Pb}$ is notably lower compared to those for other lead isotopes, which may indicate a potential issue with this particular measurement. Furthermore, the INCL-ABLA++ calculations show better agreement with the other $^{\text{nat}}\mathrm{Pb}$ datasets. For $^{207,206,204}\mathrm{Pb}$, further independent measurements are required to establish the trend of the fission cross sections at high kinetic energies. In particular, the onset of the fission cross-section plateau is observed at 400~MeV/nucleon for $^{207}\mathrm{Pb}$, a feature not present in $^{206,208}\mathrm{Pb}$ or $^{nat}\mathrm{Pb}$. Therefore, the reliability of the current $^{207}\mathrm{Pb}$ data remains questionable and requires validation.

Finally, Fig.~\ref{fig7} displays the experimental data for proton-induced fission reactions on actinides, in particular for $^{232}\mathrm{Th}$, $^{238,235,233}\mathrm{U}$, $^{237}\mathrm{Np}$, and $^{239}\mathrm{Pu}$ nuclei~\cite{EXFORWEB,Bernas2003,Kotov2006,Schmidt13,Antia_thesis}, measured in both direct and inverse kinematics. The INCL-ABLA++ model provides a good description of the existing datasets, although it slightly underestimates the $^{233}\mathrm{U}$ dataset obtained by Kotov and collaborators. In this case, additional independent measurements for $^{233}\mathrm{U}$ and $^{239}\mathrm{Pu}$ are necessary to establish the trend of the fission cross sections.

The results obtained for neutron-induced fission reactions on pre-actinides~\cite{EXFORWEB,IDEPJW146,Tarrio2011} and actinides~\cite{EXFORWEB,IDEPJW146,Tarrio2023,Manna2025} are presented in Figs.~\ref{fig8} and~\ref{fig9}, respectively. INCL-ABLA++ provides a reasonable agreement with the existing datasets for $^{\text{nat}}\mathrm{W}$, $^{197}\mathrm{Au}$, $^{\text{nat}}\mathrm{Pb}$, and $^{209}\mathrm{Bi}$ nuclei, although it slightly overestimates the data points for $^{181}\mathrm{Ta}$ and $^{205}\mathrm{Tl}$. For all nuclei below $^{\text{nat}}\mathrm{Pb}$, more independent measurements are required to establish the trend of the fission cross sections. Finally, for the actinides, the INCL-ABLA++ calculations are in good agreement with both the empirical data and the data evaluation performed by Duran and collaborators~\cite{IDEPJW146}. However, it is important to emphasize that the calculations slightly overestimate the fission cross sections for $^{232}\mathrm{Th}$ around 600~MeV/nucleon, where other nuclei exhibit a bump at that energy. Additionally, the recent measurement performed by the $n\_TOF$ collaboration for $^{235}\mathrm{U}$~\cite{Manna2025} exhibits large fluctuations between 200 and 400~MeV/nucleon, which should be addressed in future measurements.

\begin{figure}[t!]
\centering
\subfigure{\includegraphics[width=0.49\textwidth,keepaspectratio]{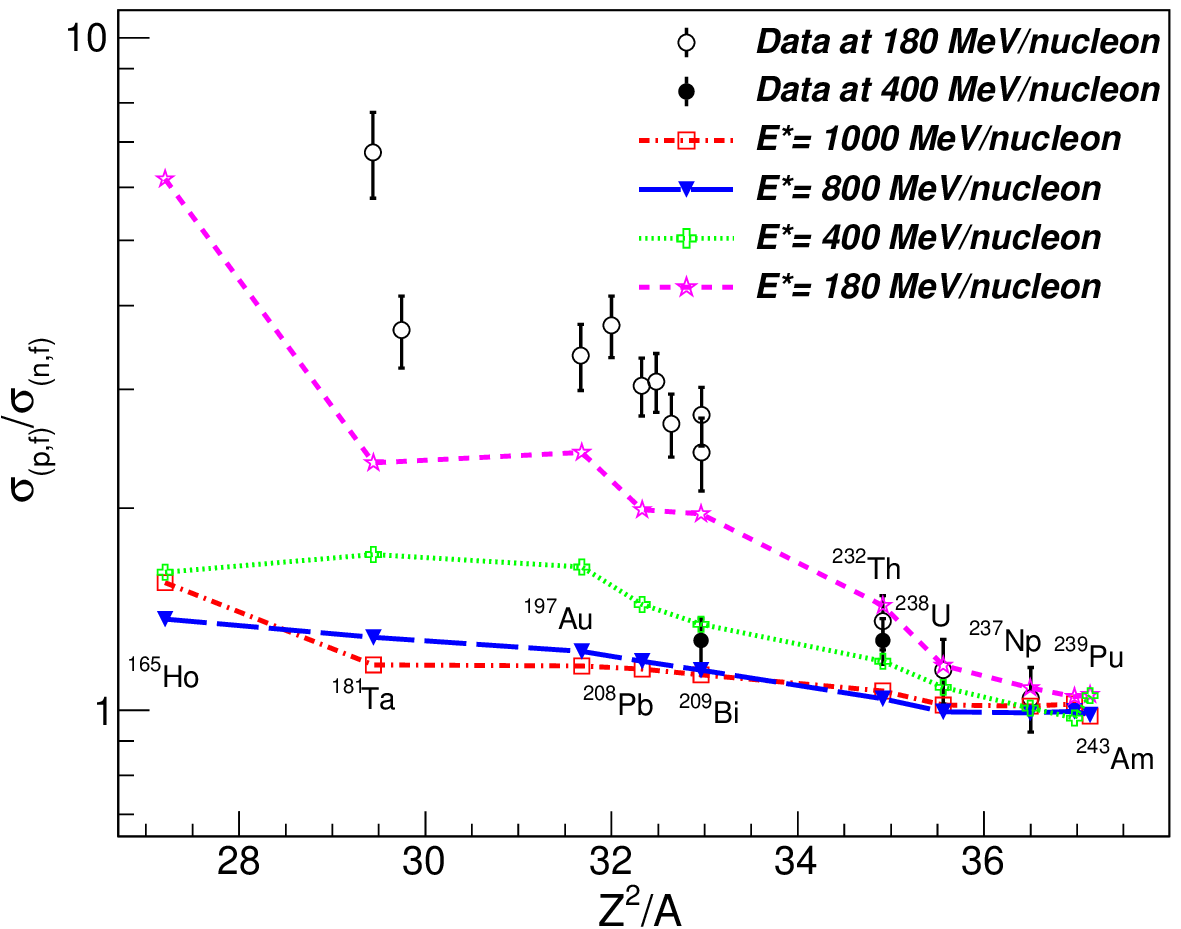}}
\caption{(Color online) Total fission cross section ratio $(p,f)$/$(n,f)$ as a function of the fissility parameter $Z^2/A$ for kinetic energies of 180~\cite{Smirnov2002} (open circles) and 400 (solid circles)~MeV/nucleon. INCL-ABLA++ calculations (lines) are also shown for different incident energies.
}
\label{fig10}
\end{figure}

The good agreement observed with the existing total fission cross sections for neutron- and proton-induced fission reactions highlights the predictive capability of the INCL-ABLA++ models. However, additional data are needed to definitively establish the trend of the fission cross sections for certain nuclei. Moreover, another open question that is not yet fully understood is the isospin dependence between the $(n,f)$ and $(p,f)$ reactions~\cite{Smirnov2002}. This dependence is illustrated in the Fig.~\ref{fig10} for different incident energies covering a large range in nuclei, from $^{165}\mathrm{Ho}$ to $^{243}\mathrm{Am}$.

The ratio $(p,f)/(n,f)$ approaches the unity for the heaviest actinides (open and solid circles), which can be attributed to the fact that nuclei in this region have very similar fission-barrier heights, resulting in minimal sensitivity to the entrance channel. However, this ratio differs significantly in the pre-actinide region, where experimental $(p,f)$ cross sections are substantially larger. This behavior is also reflected in the INCL-ABLA++ calculations at 180~MeV/nucleon (dashed line), which show similar fissility-dependent jumps. Moreover, the isospin dependence diminishes with increasing incident kinetic energy, as the ratio $(p,f)/(n,f)$ converges closer to unity at higher energies for all the nuclei. A similar trend is evident in the experimental data, where, at 400~MeV/nucleon (solid circles), the ratios $(p,f)/(n,f)$ for $^{209}\mathrm{Bi}$ and $^{232}\mathrm{Th}$ are significantly reduced and approach unity. This finding means that at high kinetic energies, the compound nuclei produced after the intranuclear cascade stage are more comparable for both $(n,f)$ and $(p,f)$ reactions. In contrast, at lower energies, the average $Z^2_{CN}/A_{CN}$ ratio is higher for proton-induced reactions leading to an increased fission probability and, consequently, higher fission cross sections compared to neutron-induced fission, as pointed out by Cond\'e and collaborators~\cite{Conde98}. To deeper understand this isospin dependence, additional independent measurements are required in the pre-actinide region, where this dependence is more pronounced, and at high kinetic energies between 200 and 1000~MeV/nucleon.
\begin{figure}[b!]
\centering
\subfigure{\includegraphics[width=0.49\textwidth,keepaspectratio]{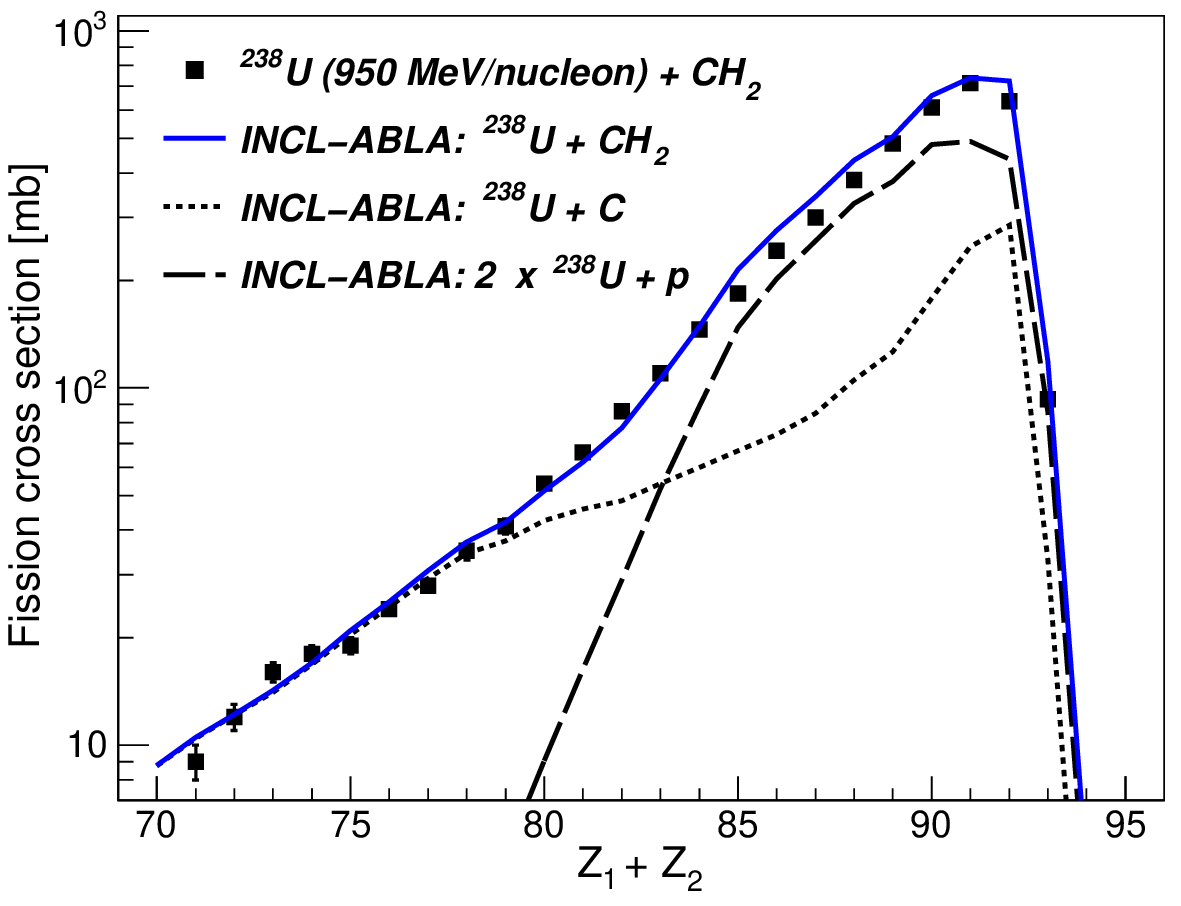}}
\caption{(Color online) Fission cross-section as a function of the atomic number of the fissioning system ($Z_1 + Z_2$) for the reaction $^{238}\mathrm{U} + \mathrm{CH_2}$ at 950~MeV/nucleon, obtained in inverse kinematics~\cite{BJPRL93}. The lines represent INCL-ABLA++ calculations for the indicated reaction (solid line), as well as for the two contributions: $^{238}\mathrm{U} + p$ (long-dashed line) and $^{238}\mathrm{U} + \mathrm{C}$ (dotted line).
}
\label{fig11}
\end{figure}

Another interesting fission observable to benchmark the INCL-ABLA++ calculations is the fission cross section as a function of the atomic number of the fissioning systems obtained as $Z_1 + Z_2$, where $Z_1$ and $ Z_2$ represent the atomic number of the final fission fragments. This observable is obtained in fission experiments performed in inverse kinematics, since this technique allows for the identification, in terms of charge, of both fission fragments with high resolution~\cite{audrey99_2019,audrey102_2020}. In Fig.~\ref{fig11}, this observable is displayed for the reaction $^{238}\mathrm{U} + \mathrm{CH_2}$ at 950~MeV/nucleon (solid squares)~\cite{BJPRL93}, which can also be simulated with the INCL-ABLA++ model. As shown in the figure, the INCL-ABLA++ calculation (solid line) is in excellent agreement with the experimental data of $^{238}\mathrm{U} + \mathrm{CH_2}$, which is the sum of two contributions: (i) proton-induced fission of $^{238}\mathrm{U}$~\footnote{The factor of 2 accounts for the two hydrogen atoms in the $\mathrm{CH_2}$ molecule.} (long-dashed line) and (ii) carbon-induced fission of $^{238}\mathrm{U}$ (dotted line). Moreover, this observable is anti-correlated with the excitation energy, as lighter fissioning systems are produced in more central collisions~\cite{CSPRL07,schmitt_schmidt}, which induce higher excitation energy in the compound nuclei formed after spallation and fragmentation reactions, as illustrated in Fig.~\ref{fig1}. Therefore, the good agreement with this dataset, which covers a wide range of fissioning systems from $Z_1 + Z_2=71$ to 93, demonstrates that INCL-ABLA++ provides a reasonable description of the induced excitation energy, as well as the subsequent de-excitation through particle emission and fission.
\begin{figure}[b!]
\centering
\subfigure{\includegraphics[width=0.49\textwidth,keepaspectratio]{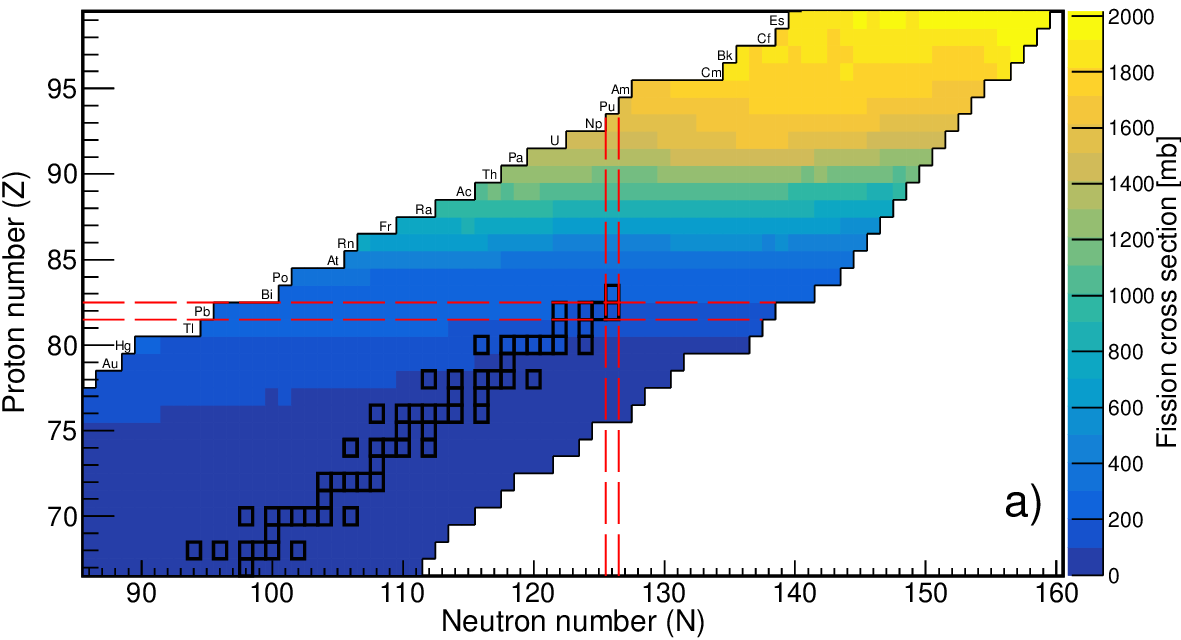}\label{fig12a}}
\centering
\subfigure{\includegraphics[width=0.49\textwidth,keepaspectratio]{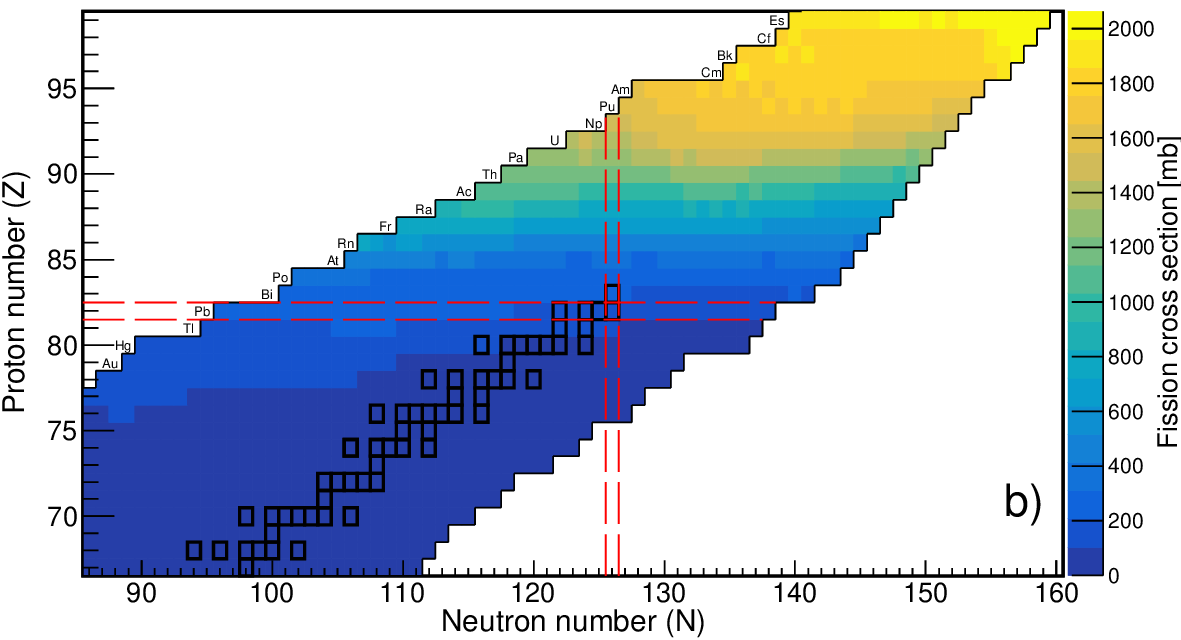}\label{fig12b}}
\centering
\subfigure{\includegraphics[width=0.47\textwidth,keepaspectratio]{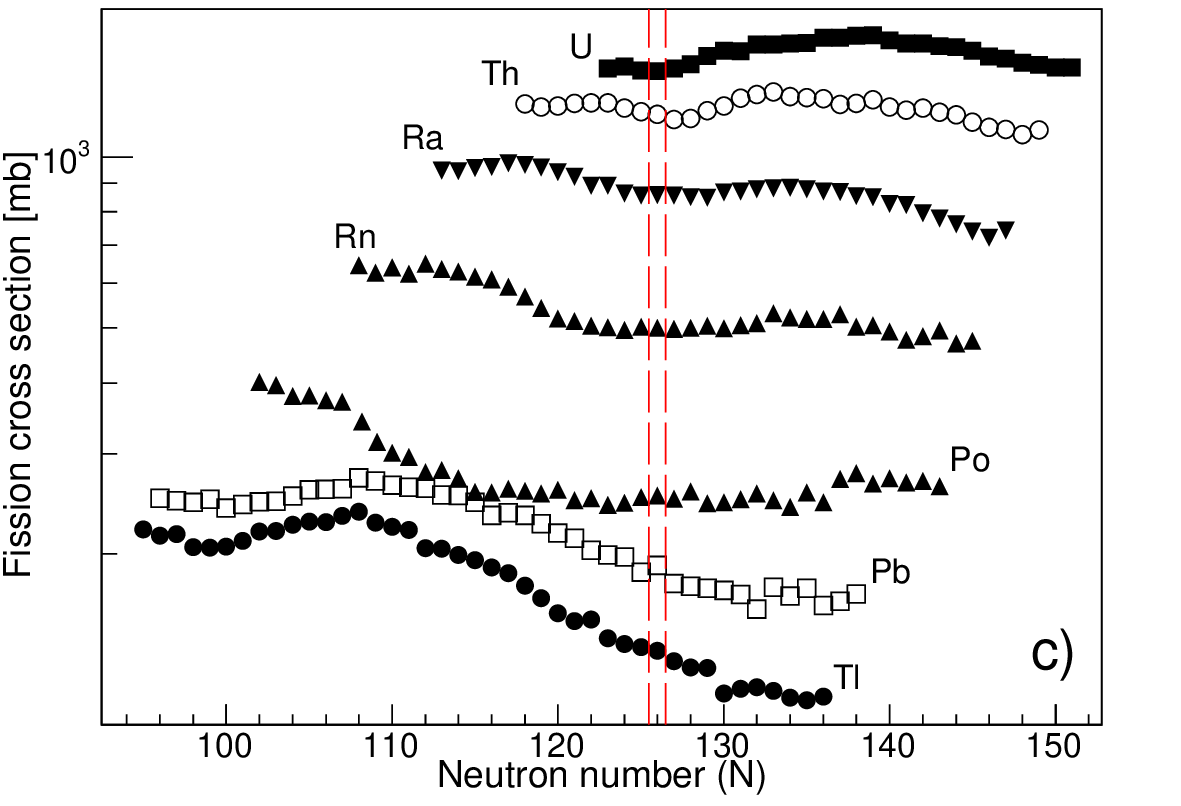}\label{fig12c}}
\caption{(Color online) Chart of (a) proton- and (b) neutron-induced total fission cross sections as a function of the target nucleus proton ($Z$) and neutron ($N$) numbers at an incident kinetic energy of 1~GeV/nucleon. For orientation, the primordial stable isotopes are marked by black open squares. The boundaries of known nuclei were determined using the atomic mass evaluation AME2020~\cite{mass2020}. The pad (c) shows the evolution of fission cross sections induced by protons with the neutron number ($N$) for different elements. The horizontal (vertical) lines correspond to the proton~(neutron) shell $Z=82$~($N=126$).
}
\label{fig12}
\end{figure}

Finally, the dynamical INCL-ABLA++ approach is used to investigate total fission cross sections of exotic nuclei, far from the valley of stability, induced by protons and neutrons, which is illustrated in Figs.~\ref{fig12a} and~\ref{fig12b} at an incident energy of 1~GeV/nucleon, respectively. In these figures, the total fission cross section is calculated for exotic nuclei between the limits defined by the atomic mass evaluation AME2020~\cite{mass2020}, as well as for stable nuclei (black open squares). In the actinide region, the cross sections are observed to be approximately 100 times larger than those in the pre-actinide region, as shown in the previous figures. This difference is clearly attributed to the transition from low fission barriers in the actinide region~($\approx$~4~MeV) to high fission barriers in the pre-actinides region~($\approx$~25~MeV). Additionally, it is observed that the fission cross section decreases with increasing the neutron number within a given isotopic chain, as shown in Fig.~\ref{fig12c} for different elements. This can be attributed to the fact that spallation reactions on more neutron-rich nuclei populate compound nuclei with higher fission-barrier heights (see Fig.~\ref{fig3}), leading to lower fission cross sections. Additionally, this effect is enhanced by the lower neutron separation energies of neutron-rich nuclei, which favor the neutron evaporation de-excitation channel, thereby increasing the probability of producing evaporation residues. In the region of actinides around $N=126$, the fission cross sections decrease due to the observed bump in the fission-barrier heights. This effect is enhanced by the fact that, in the actinide region, fission reactions predominantly occur at low excitation energies, where the fission-barrier height plays a significant role. For the neutron-deficient side of pre-actinides, such as Tl and Pb, the fission cross sections decrease due to competition with proton emission, which is favored because of the low proton separation energies.

\section{Conclusions}
The dynamical Liège intranuclear cascade model (INCL++), integrated with the ABLA++ de-excitation model, has been benchmarked against several datasets of total fission cross sections accumulated over recent decades. These datasets encompass neutron- and proton-induced fission reactions and span a broad range of target nuclei, from pre-actinides to actinides, as well as incident energies from a few MeVs/nucleon up to 1~GeV/nucleon.

To improve the model benchmark, the description of fission-barrier heights was refined using empirical data, particularly improving the even-odd effects in the actinide region. The calculation of particle separation energies has also been updated to the atomic mass evaluation AME2020~\cite{mass2020}. In addition, the excitation-energy-dependent dissipation coefficient, originally included in the legacy Fortran-based ABLA07 code, has been removed to achieve more accurate agreement with fission cross sections at high excitation energies, a modification that was also done in previous publications with ABLA07~\cite{Yassid89_2014,JLRS14,YAPRC91,JLRS15_1,JLRS16_2,JLRS16_1,JLRS15_2}.

Total fission cross sections for the reactions n + $^{\text{nat}}\mathrm{Pb}$ and p + $^{\text{208}}\mathrm{Pb}$, as a function of the incident kinetic energy, have been simultaneously analyzed to investigate transient time effects in fission, revealing identical behavior for both reactions. This demonstrates that the transient time effects are important in both cases and are independent of the entrance channel.

The resulting predictions for total fission cross sections induced by neutrons and protons show good agreement with the available experimental data, underscoring the models' applicability and predictive capability for incident energies above 20~MeV/nucleon for pre-actinides and 150~MeV/nucleon for actinides. However, additional data on fission cross sections in the energy range of 100~MeV/nucleon to 1~GeV/nucleon are essential to further refine the model parameters. Moreover, measurements of fission-barrier heights for exotic nuclei far from the stability valley are also important for improving the model's accuracy and gaining deeper insights into the isospin dependence of total fission cross sections, as well as its evolution with the neutron number within a given isotopic chain.

Finally, the enhanced INCL-ABLA++ framework has been employed to explore the fission cross sections of exotic heavy fissile nuclei for masses $A > 155$ and near the limits of nuclear stability, as defined by the atomic mass evaluation AME2020~\cite{mass2020}. These exotic nuclei may be produced through fragmentation reactions at next-generation facilities, such as HELIOS at Argonne National Laboratory~\cite{Helios,Bennett2023}, the ISOLDE Solenoidal Spectrometer~(ISS) at CERN~\cite{Isolde}, SOLARIS at Facility for Rare Isotope Beams (FRIB)~\cite{Frib}, and R$^3$B at GSI/FAIR~\cite{Aumann2024}. In particular, the advanced R$^3$B/SOFIA~\cite{JLRS14,JLRS15_1,audrey99_2019,audrey102_2020} fission setup is expected to provide some of the required data for proton-induced fission reactions~\cite{AGG_epj2023,GGJ_epj2023,JLRS_epj2023} in the next decade. Moreover, the development of new parameter optimization algorithms based on bayesian generalized least squares methods will enable a more comprehensive study of the uncertainties in the INCL and ABLA++ parameters~\cite{Hirtz2024}, offering improved constraints on these models.
\begin{acknowledgments}
We thank the support by the NURBS project (Grant No. 219157) and from the “María de Maeztu” Grants MDM-2016-0692 and CEX2023-001318-M, funded by MICIU/AEI/10.13039/501100011033. A.G.-G and G.G.-J. thank the support by the Spanish Ministry of Innovation and Science under Grants PRE2018-085934 and PRE2019-087415, respectively. J.L.R.-S. is thankful for the support by Xunta de Galicia under the programme “Proyectos de excelencia” Grant No. ED431F-2023/43, and by the "Ram\'{o}n y Cajal" programme under Grant No. RYC2021-031989-I funded by MCIN/AEI/10.13039/501100011033 and by “European Union NextGenerationEU/PRTR”.
\end{acknowledgments}

\bibliographystyle{unsrt}

\end{document}